\documentclass[nojss]{jss}

\author{Mark D. Risser\\The Ohio State University \And 
        Catherine A. Calder\\The Ohio State University}
\title{Local Likelihood Estimation for Covariance Functions with Spatially-Varying Parameters: The \pkg{convoSPAT} Package for \proglang{R}}

\Plainauthor{Mark D. Risser, Catherine A. Calder} 
\Plaintitle{Local Likelihood Estimation for Covariance Functions with Spatially-Varying Parameters: The convoSPAT Package for R} 
\Shorttitle{\pkg{convoSPAT}: Local Likelihood Estimation for Spatially-Varying Covariance Functions} 

\Abstract{
In spite of the interest in and appeal of convolution-based approaches for nonstationary spatial modeling, off-the-shelf software for model fitting does not as of yet exist. Convolution-based models are highly flexible yet notoriously difficult to fit, even with relatively small data sets. The general lack of pre-packaged options for model fitting makes it difficult to compare new methodology in nonstationary modeling with other existing methods, and as a result most new models are simply compared to stationary models. Using a convolution-based approach, we present a new nonstationary covariance function for spatial Gaussian process models that allows for efficient computing in two ways: first, by representing the spatially-varying parameters via a discrete mixture or ``mixture component'' model, and second, by estimating the mixture component parameters through a local likelihood approach. In order to make computations for a convolution-based nonstationary spatial model readily available, this paper also presents and describes the \pkg{convoSPAT} package for \proglang{R}. The nonstationary model is fit to both a synthetic data set and a real data application involving annual precipitation to demonstrate the capabilities of the package.
}

\Plainkeywords{spatial statistics, nonstationary modeling, local likelihood estimation, precipitation, R} 

\Address{
  Catherine A. Calder\\
  Department of Statistics\\
  The Ohio State University\\
  1958 Neil Avenue\\
  Columbus, OH, USA 43210\\
  E-mail: \email{calder@stat.osu.edu}
}

\usepackage{amsmath, amsthm}
\usepackage{amssymb}

\newcommand{\bfbeta}{{\boldsymbol \beta}}
\newcommand{\bftheta}{{\boldsymbol \theta}}
\newcommand{\bfs}{{\bf s}}

\DeclareMathOperator{\Cov}{Cov}

\begin{document}

\section{Introduction} \label{introduction}

The Gaussian process is an extremely popular modeling approach in modern-day spatial and environmental statistics, due largely to the fact that the model is completely characterized by first- and second-order properties, and the second-order properties are straightforward to specify through widely used classes of valid covariance functions. A broad literature on covariance function modeling exists, but traditional approaches are mostly based on assumptions of isotropy or stationarity, in which the covariance between the spatial process at two locations is a function of only the separation distance or separation vector, respectively. This modeling assumption is made mostly for convenience, and is rarely a realistic assumption in practice. As a result, a wide variety of nonstationary covariance function models for Gaussian process models have been developed (e.g., \citealp{sampGutt}; \citealp{Higdon98}; \citealp{damian2001}; \citealp{fuentes2001};  \citealp{schmidt03}; \citealp{PacScher}; \citealp{calder08}; \citealp{schmidt11}; \citealp{reich2011}; and \citealp{ViannaNeto}), in which the spatial dependence structure is allowed to vary over the spatial region of interest. However, while these nonstationary approaches more appropriately model the covariance in the spatial process, most are also highly complex and require intricate model-fitting algorithms, making it very difficult to replicate their results in a general setting. Therefore, when new nonstationary methods are developed, their performance is usually compared to stationary models, for which robust software is available. While software exists for several of these nonstationary approaches (see below), there are currently no pre-packaged options for fitting convolution-based nonstationary models.

To address this need, we present a simplified version of the nonstationary spatial Gaussian process model introduced by \cite{PacScher} in which the locally-varying geometric anisotropies are modeled using a ``mixture component'' approach, similar to the discrete mixture kernel convolution approach in \cite{Higdon98} but allowing the underlying correlation structure to be specified by the modeler. The model is extended to allow other properties to vary over space as well, such as the process variance, nugget effect, and smoothness. An additional degree of efficiency is gained by using local likelihood techniques to estimate the spatially-varying features of the spatial process; then, the locally estimated features are smoothed over space, similar in nature to the approach of \cite{fuentes02}. 

This paper also presents and describes the \pkg{convoSPAT} package for \proglang{R} for conducting a full analysis of point-referenced spatial data using a convolution-based nonstationary spatial Gaussian process model. The primary contribution of the package is to provide accessible model-fitting tools for spatial data of this type, as software for convolution-based nonstationary modeling does not currently exist. Furthermore, the methods used by the package are computationally efficient even when the size of the data is relatively large (on the order of $n=1000$). The package is able to handle both a single realization of the spatial process observed at a finite set of locations, as well as independent and identically distributed replicates of the spatial process observed at a common set of locations. Finally, the paper demonstrates how the package can be used, and provides analyses of both simulated and real data sets.

As noted, there are several other (albeit non convolution-based) methods for nonstationary spatial modeling that do offer software, namely the basis function approach in the \pkg{fields} package (\citealp{R_fields}) and the Gaussian Markov random field approach in the \pkg{INLA} package (\citealp{Lindgren2011}; \citealp{spde13}; \citealp{Fuglstad2015}; \citealp{Lindgren2015}), both available for \proglang{R}. However, as these methods arise from significantly different modeling approaches, the \pkg{convoSPAT} package represents a novel contribution to the set of available software for nonstationary spatial modeling. Comparison across software for these various packages is beyond the scope of this paper and will be reserved for future work.

The paper is organized as follows. Section \ref{sec2} introduces a convolution-based approach for nonstationary spatial statistical modeling, and Section \ref{NSGPM} describes a full model for observed data and the mixture component parameterization. Section \ref{CEI} outlines a computationally efficient approach to inference for the model introduced in Section \ref{NSGPM}; Sections \ref{sec5}, \ref{simulateddataeg}, and \ref{USprecip} outline usage of the \pkg{convoSPAT} package and present two applications. Section \ref{discussion} concludes the paper.

\section{A convolution-based nonstationary covariance function} \label{sec2}

Process convolutions or moving average models are popular constructive methods for specifying a nonstationary process model. In general, a spatial stochastic process $Y(\cdot)$ on $G \subset \mathcal{R}^d$ can be defined by the kernel convolution 
\begin{equation} \label{kernelconvolution}
Y({\bf s}) = \int_{\mathcal{R}^d} K_{\bf s}({\bf u}) dW({\bf u}),
\end{equation}
where $W(\cdot)$ is a $d$-dimensional stochastic process and $K_{\bf s}(\cdot)$ is a (possibly parametric) spatially-varying kernel function centered at ${\bf s} \in G$. \cite{higdon2} summarizes the extremely flexible class of spatial process models defined by Equation~\ref{kernelconvolution}: see, for example, \cite{Barry1996}, \cite{VerHoef2004}, \cite{Wolpert1999}, \cite{Higdon98}, \cite{VerHoef2004}, and \cite{Barry_VerHoef1998}. 

The kernel convolution Equation~\ref{kernelconvolution} defines a mean-zero nonstationary spatial Gaussian process (GP) if $W(\cdot)$ is chosen to be $d$-dimensional Brownian motion. A benefit of using Equation~\ref{kernelconvolution} is that in this case the kernel functions completely specify the second-order properties of the GP through the covariance function
\begin{equation} \label{covfcn}
\Cov( Y({\bf s}), Y({\bf s')} ) = E \big[Y({\bf s})Y({\bf s'}) \big] = \int_{G}  K_{\bf s}({\bf u}) K_{\bf s'}({\bf u}) d{\bf u},
\end{equation}
where ${\bf s, s'} \in G$. The popularity of this approach is due largely to the fact that it is much easier to specify kernel functions than a covariance function directly, since the kernel functions only require
$
\int_{\mathcal{R}^d} K_{\bf s}({\bf u}) d{\bf u} <\infty
$
and
$
\int_{\mathcal{R}^d} K^2_{\bf s}({\bf u}) d{\bf u} <\infty,
$
while a covariance function must be even and nonnegative definite (\citealp{BochnerBook}; \citealp{AdlerBook}). A famous result (\citealp{thiebaux76}; \citealp{thiebaux_pedder}) uses a parametric class of Gaussian kernel functions in Equation~\ref{covfcn} to give a closed-form covariance function; this result was later extended (\citealp{paciorek2003}; \citealp{PacScher}; \citealp{stein2005}) to show that
\begin{equation} \label{NScov}
C^{NS}({\bf s}, {\bf s'}; \bftheta) = \sigma({\bf s}) \sigma({\bf s'}) \frac{\left|{\bf \Sigma(s)}\right|^{1/4} \left|{\bf \Sigma(s')}\right|^{1/4} }{ \left|\frac{{\bf \Sigma(s) + \Sigma(s')}}{2}\right|^{1/2} } g \left( \sqrt{Q({\bf s, s'})} \right),
\end{equation}
is a valid, nonstationary, parametric covariance function on $\mathcal{R}^d, d \geq 1$, when $g(\cdot)$ is chosen to be a valid correlation function on $\mathcal{R}^d, d \geq 1$. Note that Equation~\ref{NScov} no longer requires kernel functions to be specified. In Equation~\ref{NScov}, $\boldsymbol{\theta}$ is a generic parameter vector, $\sigma(\cdot)$ represents a spatially-varying standard deviation, ${\bf \Sigma}(\cdot)$ is a $d\times d$ matrix that represents the spatially-varying local anisotropy (controlling both the range and direction of dependence), and
\begin{equation} \label{Q}
Q({\bf s, s'}) = {\bf (s-s')}^\top \left(\frac{{\bf \Sigma(s) + \Sigma(s')}}{2}\right)^{-1}{\bf (s-s')}
\end{equation}
is a Mahalanobis distance. Furthermore, choosing $g(\cdot)$ to be the Mat\'{e}rn correlation function also allows for the introduction of $\kappa(\cdot)$, a spatially-varying smoothness parameter (\citealp{stein2005}; in this case, the Mat\'{e}rn correlation function in Equation~\ref{NScov} has smoothness $[\kappa({\bf s}) + \kappa({\bf s}')]/2$). While using Equation~\ref{NScov} no longer requires the notion of kernel convolution, we refer to ${\bf \Sigma}(\cdot)$ as the kernel matrix, since it was originally defined as the covariance matrix of a Gaussian kernel function (\citealp{thiebaux76}; \citealp{thiebaux_pedder}). The covariance function in Equation~\ref{NScov} is extremely flexible, and has been used in various forms throughout the literature, e.g., \cite{PacScher}, \cite{Anderes_Stein}, \cite{kleiber2012}, \cite{Katzfuss2013}, and \cite{Risser15}. 

\section{A nonstationary spatial Gaussian process model} \label{NSGPM}

The covariance function in Equation~\ref{NScov} can be used to define a nonstationary spatial Gaussian process model using the following framework. Let $\{ Z({\bf s}), {\bf s} \in G \}$ be a spatial field defined on $G \subset \mathcal{R}^d$, where
\begin{equation} \label{model}
Z({\bf s}) = {\bf x}({\bf s})^\top \boldsymbol{\beta} + Y({\bf s}) + \epsilon({\bf s}).
\end{equation}
In Equation~\ref{model}, the mean of the spatial field is $E[Z({\bf s})] = {\bf x}({\bf s})^\top \boldsymbol{\beta}$, where ${\bf x}({\bf s})$ is a $p$-vector of covariates for location ${\bf s}$ and $\boldsymbol{\beta} \in \mathcal{R}^p$ are unknown regression coefficients. $Y(\cdot)$ represents a spatially-dependent, mean-zero Gaussian process with covariance function $C^{NS}$ from Equation~\ref{NScov}, while $\epsilon(\cdot)$ represents measurement error and, given $\tau^2(\cdot)$, is conditionally independent $\mathcal{N}(0, \tau^2({\bf s}))$. (Note: $\mathcal{N}(a, b)$ denotes the univariate Gaussian distribution with mean $a$ and variance $b$.) The spatially-referenced random components, $\epsilon(\cdot)$ and $Y(\cdot)$, are assumed to be mutually independent. Finally, define $\bftheta$ to be a vector of all the variance-covariance parameters from the Gaussian process $Y(\cdot)$ and error process $\epsilon(\cdot)$. 

Now, suppose we have observations which are a partial realization of $Z(\cdot)$, taken at a fixed, finite set of $n$ spatial locations $\{{\bf s}_i\}_{i=1}^n \in G$, giving the random (observed) vector ${\bf Z} = \left( Z({\bf s}_1), ... , Z({\bf s}_n) \right)$. The model in Equation~\ref{model} implies that ${\bf Z}$ has a multivariate Gaussian distribution, conditional on the unobserved latent process ${\bf Y} = \left( Y({\bf s}_1), ... , Y({\bf s}_n) \right)$ and all other model parameters:
\begin{equation} \label{ZYdist}
{\bf Z} | {\bf Y}, \bfbeta, \bftheta \sim \mathcal{N}_n\big({\bf X}\boldsymbol{\beta} + {\bf Y}, {\bf D}(\bftheta) \big),
\end{equation}
where the $i$th row of ${\bf X}$ is ${\bf x}({\bf s}_i)$ and ${\bf D}(\bftheta)$ is a diagonal matrix with $(i, i)$ element $\tau^2(\bfs_i)$. (Note: $\mathcal{N}_q({\bf a}, {\bf B})$ denotes the $q$-variate Gaussian distribution with mean vector ${\bf a}$ and covariance ${\bf B}$.) Integrating out the process ${\bf Y}$ from Equation~\ref{ZYdist}, we can obtain the marginal likelihood of the observed data ${\bf Z}$ given all parameters, which is 
\begin{equation} \label{Zdist}
{\bf Z} | \bfbeta, \bftheta \sim \mathcal{N}_n\big({\bf X}\boldsymbol{\beta}, {\bf D(\bftheta) + \Omega(\bftheta)} \big),
\end{equation}
where  ${\bf \Omega(\bftheta)}$ has elements $\Omega_{ij}(\bftheta) = C^{NS}({\bf s}_i, {\bf s}_j; \boldsymbol{\theta})$. For a particular application, the practitioner can specify the underlying correlation structure (through $g(\cdot)$) as well as determine which of $\{ {\bf \Sigma}(\cdot), \sigma(\cdot), \tau^2(\cdot) \}$ (or $ \kappa(\cdot)$, if the Mat\'{e}rn is used) should be fixed or allowed to vary spatially. However, some care should be taken in choosing which quantities should be spatially-varying: for example, \cite{Anderes_Stein} note that allowing both ${\bf \Sigma}(\cdot)$ and $\kappa(\cdot)$ to vary over space leads to issues with identifiability.

\subsection{Discrete mixture representation} \label{DMR}

One way to reduce the computational demands of fitting a Gaussian process-based spatial model with parametric covariance function given by Equation~\ref{NScov} is by characterizing the nonstationary behavior of a spatial process through the discretized basis kernel approach of \cite{Higdon98}. \cite{Higdon98} estimated the Gaussian kernel function for a generic location to be a weighted average of ``basis'' kernel functions, estimated locally over the spatial region of interest. However, since the use of Gaussian kernel functions results in undesirable smoothness properties (see, e.g., \citealp{PacScher}), we instead use a related ``mixture component'' approach, in which the parametric quantities for an arbitrary spatial location are defined as a mixture of spatially-varying parameter values associated with a fixed set of component locations. Specifically, in this new approach, we define mixture component locations $\{ {\bf b}_k : k = 1, \dots, K\}$ with corresponding parameters $\{ ({\bf \Sigma}_k, \sigma^2_k, \tau^2_k, \kappa_k) : k = 1, \dots, K \}$ (which are the kernel matrix, variance, nugget variance, and smoothness, respectively). Then, for $\phi \in \{ {\bf \Sigma}, \sigma^2, \tau^2, \kappa \}$, the parameter set for an arbitrary location ${\bf s} \in G$ is calculated as
\begin{equation} \label{spat_var_par}
\phi({\bf s}) = \sum_{k = 1}^K w_k({\bf s}) \phi_k,
\end{equation}
where
\begin{equation} \label{smoother}
w_k({\bf s}) \propto \exp \left\{ -\frac{|| {\bf s - b}_k || ^2}{2\lambda_w}  \right\}
\end{equation}
such that $\sum_{k=1}^K w_k({\bf s}) = 1$. For example, the kernel matrix for ${\bf s} \in G$ is ${\bf \Sigma(s)} = \sum_{k = 1}^K w_k({\bf s}) {\bf \Sigma}_k$. In Equation~\ref{smoother}, $\lambda_w$ acts as a tuning parameter, ensuring that the rate of decay in the weighting function is appropriate for both the data set and scale of the spatial domain. Using this approach, the number of parameters is now linear in $K$, the number of mixture component locations, instead of $n$, the sample size. Furthermore, this specification still enables the modeler to choose which parameters should be spatially-varying: the kernel matrices, the process variance, the nugget variance, and the smoothness.

\subsection{Prediction} \label{predic}

Define ${\bf Z}^* = \left( Z({\bf s}^*_1), ... , Z({\bf s}^*_m) \right)$ to be a vector of the process values at all prediction locations of interest. The Gaussian process model in Equation~\ref{model} implies that
\[
\left[ \begin{array}{c} {\bf Z} \\ {\bf Z^*} \end{array} \Bigg| \hskip1ex \boldsymbol{\beta, \theta} \hskip1ex \right] \sim \mathcal{N}_{n+m} \left( \left[ \begin{array}{c} {\bf X}\boldsymbol{\beta} \\ {\bf X}^*\boldsymbol{\beta} \end{array} \right] ,  \left[ \begin{array}{cc} {\bf D}(\bftheta) + {\bf \Omega}(\bftheta) & {\bf \Omega_{ZZ^*}(\bftheta)} \\  {\bf \Omega_{Z^*Z}(\bftheta)} &  {\bf D}^*(\bftheta) + {\bf \Omega}^*(\bftheta)  \end{array} \right] \right),
\]
where $\Cov({\bf Z}^*) = {\bf D}^*(\bftheta) +  {\bf \Omega}^*(\bftheta)$ and $\Cov({\bf Z,  Z}^*) = {\bf \Omega_{ZZ^*}(\bftheta)}$. By the properties of the multivariate Gaussian distribution,
\begin{equation} \label{Zstar}
{\bf Z^* | Z = z}, \boldsymbol{\beta, \theta} \sim \mathcal{N}_m (\boldsymbol{\mu}_{\bf Z^*|z} , {\bf \Sigma}_{\bf Z^*|z} ),
\end{equation}
where
\begin{equation} \label{kriging1}
\boldsymbol{\mu}_{\bf Z^*|z} =  {\bf X}^*\boldsymbol{\beta} + {\bf \Omega_{Z^*Z}}(\bftheta) [{\bf D}(\bftheta) + {\bf \Omega}(\bftheta)]^{-1} ({\bf z} - {\bf X}\boldsymbol{\beta}),
\end{equation}
and
\begin{equation} \label{kriging2}
{\bf \Sigma}_{\bf Z^*|z} = [ {\bf D}^*(\bftheta) + {\bf \Omega}^*(\bftheta) ] - {\bf \Omega_{Z^*Z}}(\bftheta) [{\bf D}(\bftheta) + {\bf \Omega}(\bftheta)]^{-1}  {\bf \Omega_{ZZ^*}(\bftheta)}. 
\end{equation}
Using plug-in estimates $\widehat{\boldsymbol{\beta}}$ and $\widehat{\boldsymbol{\theta}}$ (see Section \ref{CEI}), the predictor for ${\bf Z}^*$ is then $\widehat{\boldsymbol{\mu}}_{\bf Z^*|z}$ with corresponding prediction errors as the square root of the diagonal elements of $\widehat{\bf \Sigma}_{\bf Z^*|z}$.

\subsubsection{Out-of-sample evaluation criteria}

Three cross-validation evaluation criteria can be used to assess the fit of the nonstationary spatial model given in Equation~\ref{model}. First, the mean squared prediction error
\begin{equation} \label{MSPE}
MSPE = \frac{1 }{ m  }  \sum_{j=1}^m (z^*_j - \widehat{z}^*_j)^2,
\end{equation}
where $z^*_j$ is the $j$th held-out observed (or ``validation'') value and $\widehat{z}^*_j$ is the corresponding predictor (from Equation~\ref{kriging1}). The MSPE is a point-wise measure of model fit, and smaller MSPE indicates better predictions. 

Second, to assess the prediction error relative to the standard error of each prediction, we use the so-called prediction mean squared deviation ratio
\begin{equation} \label{pMSDR}
pMSDR = \frac{1 }{ m  }  \sum_{j=1}^m \frac{(z^*_j - \widehat{z}^*_j)^2}{\widehat{\sigma}_j},
\end{equation}
where $z^*_j$ and $\widehat{z}^*_j$ are defined as above and $\widehat{\sigma}_j$ is the prediction error corresponding to $\widehat{z}^*_j$ (from Equation~\ref{kriging2}).

Finally, the continuous rank probability score will be used (a proper scoring rule; see \citealp{PropScoring}). For the $j$th prediction, this is defined as
\begin{equation} \label{CRPS}
CRPS_j \equiv CRPS(F_j, z^*_j) = - \int_{-\infty}^{\infty} \big( F_j(x) - {1}  \{ x \geq z^*_j \} \big)^2dx,
\end{equation}
where $F_j(\cdot)$ is the cumulative distribution function (CDF) for the predictive distribution of $z^*_j$ given the training data and ${1}\{ \cdot \}$ is the indicator function. In this case, given that the predictive CDF is  Gaussian (conditional on parameters; see Equation~\ref{Zstar}), a computational shortcut can be used for calculating Equation~\ref{CRPS}: when $F$ is Gaussian with mean $\mu$ and variance $\sigma^2$,
\begin{equation*} \label{CRPS_normal}
CRPS \big(  F,  z^*_j \big) = \sigma \Bigg[ \frac{1}{\sqrt{\pi}} - 2\cdot \phi\left(\frac{z^*_j - \mu}{\sigma} \right) - \frac{z^*_j - \mu}{\sigma} \left( 2 \cdot \Phi\left( \frac{z^*_j - \mu}{\sigma} \right) - 1 \right) \Bigg],
\end{equation*}
where $\phi$ and $\Phi$ denote the probability density and cumulative distribution functions, respectively, of a standard Gaussian random variable. The reported metric will be the average over all validation locations,
$
\widehat{CRPS} = m^{-1}  \sum_{j=1}^m \widehat{CRPS}_j.
$
CRPS measures the fit of the predictive density; larger CRPS (i.e., smaller negative values) indicates better model fit.

\section{Computationally efficient inference} \label{CEI}

As discussed in Section \ref{introduction}, fast and efficient inference for a nonstationary process convolution model has yet to be made readily available for general use. In spite of its popularity, the use of Equation~\ref{NScov} always requires some kind of constraints and has suffered from a lack of widespread use due to the complexity of the requisite model fitting and limited pre-packaged options. Focusing on the spatially-varying local anisotropy matrices ${\bf \Sigma}(\cdot)$, the covariance function in Equation~\ref{NScov} requires a kernel matrix at every observation and prediction location of interest. \cite{PacScher} accomplish this by modeling ${\bf \Sigma}(\cdot)$ as itself a (stationary) stochastic process, assigning Gaussian process priors to the elements of the spectral decomposition of ${\bf \Sigma}(\cdot)$; alternatively, \cite{Katzfuss2013} uses a basis function representation of ${\bf \Sigma}(\cdot)$. Both of these models are highly parameterized and require intricate Markov chain Monte Carlo methods for model fitting. 

The approach we propose provides efficiency in two ways: first, from the model itself, which uses a discrete mixture representation (see Section \ref{DMR}), and second, by fitting the mixture components of the model locally, using the idea of local likelihood estimation (\citealp{TibshiraniHastie}). 

\subsection{Local likelihood estimation}

Using the discrete mixture representation of Equation~\ref{spat_var_par}, a ``full likelihood'' approach to parameter estimation could be taken, in either a Bayesian or maximum likelihood framework, although the optimization in a maximum likelihood approach could become intractable for either moderately large $K$ or large $n$. However, since the primary goal of this new methodology is computational speed, a further degree of efficiency can be gained by using local likelihood estimation (LLE; \citealp{TibshiraniHastie}). 

Before discussing the local likelihood approach, we outline a restricted maximum likelihood (REML; see \citealp{patterson_thompson1971}, \citealp{patterson_thompson1974}, and \citealp{kitanidis1983}) approach for separating estimation of the mean parameters $\boldsymbol{\beta}$ from the covariance parameters $\boldsymbol{\theta}$. The full log-likelihood for $\boldsymbol{\beta}$ and $\boldsymbol{\theta}$ in Equation~\ref{model} is
\begin{equation} \label{full_lik}
\mathcal{L}^F(  \boldsymbol{\beta, \theta}; {\bf Z}) = -\frac{1}{2} \log |{\bf \Omega} + {\bf D} | - \frac{1}{2} ({\bf Z - X\boldsymbol{\beta}})^\top ({\bf \Omega} + {\bf D})^{-1}  ({\bf Z - X\boldsymbol{\beta}}),
\end{equation}
where we have abbreviated ${\bf D}\equiv{\bf D}(\bftheta)$ and ${\bf \Omega}\equiv{\bf \Omega}(\bftheta)$; a standard maximum likelihood approach would set out to maximize $\mathcal{L}^F(  \boldsymbol{\beta, \theta}; {\bf Z})$ directly. REML, on the other hand, uses a (log) likelihood based on $n-p$ linearly independent linear combinations of the data that have an expected value of zero for all possible $\boldsymbol{\beta}$ and $\boldsymbol{\theta}$. Regardless of which set of linearly independent combinations is chosen, the ``restricted'' log-likelihood, which depends only on $\boldsymbol{\theta}$, is
\begin{equation} \label{rest_lik}
\mathcal{L}^R(  \boldsymbol{\theta}; {\bf Z}) = -\frac{1}{2} \log |{\bf \Omega} + {\bf D}| -\frac{1}{2} \log |{\bf X}^\top({\bf \Omega} + {\bf D})^{-1} {\bf X}| - \frac{1}{2} {\bf Z}^\top {\bf P} {\bf Z},
\end{equation}
where
\begin{equation} \label{Pmat}
{\bf P} = ({\bf \Omega} + {\bf D})^{-1} - ({\bf \Omega} + {\bf D})^{-1} {\bf X} \big({\bf X}^\top({\bf \Omega} + {\bf D})^{-1} {\bf X}\big)^{-1} {\bf X}^\top ({\bf \Omega} + {\bf D})^{-1}.
\end{equation}
The REML estimate of $\boldsymbol{\theta}$ is obtained by maximizing $\mathcal{L}^R(  \boldsymbol{\theta}; {\bf Z})$, and the estimate of $\boldsymbol{\beta}$ is the generalized least squares estimate 
\begin{equation} \label{beta_GLS}
\widehat{\boldsymbol{\beta}} = \big({\bf X}^\top(\widehat{{\bf \Omega}} + \widehat{ {\bf D}})^{-1}{\bf X}\big)^{-1} {\bf X}^\top (\widehat{{\bf \Omega}} + \widehat{ {\bf D} })^{-1} {\bf Z},
\end{equation}
which is obtained by plugging in $\widehat{\boldsymbol{\theta}}$ to calculate $\widehat{{\bf \Omega}}$ and $\widehat{{\bf D}}$. These parameter estimates can then be plugged in to $\widehat{\boldsymbol{\mu}}_{\bf Z^*|z}$ and $\widehat{\bf \Sigma}_{\bf Z^*|z}$ to obtain predictions and prediction standard errors.

In the LLE approach, instead of maximizing Equation~\ref{rest_lik} directly we will set out to maximize $\mathcal{L}^R_k(  \boldsymbol{\theta}_{N_k}; {\bf Z}_{N_k})$, where $N_k \equiv N_k(r)$ is a neighborhood for each mixture component location ${\bf b}_k$ that depends on the radius $r$, such that
\[
N_k = \big\{ {\bf s}_i \in \{ {\bf s}_1, \dots, {\bf s}_n \} :  \{ || {\bf s}_i  -{\bf b}_k || \leq r \} \big\}
\]
and
\[
{\bf Z}_{N_k} =  \big\{ Z({\bf s}) : {\bf s} \in N_k \big\}.
\]
Correspondingly, $\boldsymbol{\theta}_{N_k} = ( {\bf \Sigma}_k, \sigma_k^2, \tau_k^2, \kappa_k )$. The radius $r$ defines the ``span'' (\citealp{TibshiraniHastie}) or window size for each mixture component. The restricted log-likelihood for neighborhood $N_k$ will be based on a stationary version of the spatial model in Equation~\ref{model}, namely
\begin{equation} \label{Smodel}
\widetilde{Z}({\bf s}) = {\bf x}({\bf s})^\top\widetilde{\boldsymbol{\beta}} + \widetilde{Y}({\bf s}) + \widetilde{\epsilon}({\bf s}),
\end{equation}
where $\widetilde{Y}(\cdot)$ is a stationary, mean-zero spatial process with covariance function
\begin{equation} \label{ani_model}
C^S({\bf s - s'}) = \sigma^2 g\left( || {\bf \Sigma}^{-1/2} ({\bf s - s'}) || \right),
\end{equation}
the $\widetilde{\epsilon}(\cdot)$ are independent and identically distributed as $\mathcal{N}(0, \tau^2)$, conditional on $\tau^2$, and again $\widetilde{Y}(\cdot)$ and $\widetilde{\epsilon}(\cdot)$ are independent. Again, in a REML framework, only the variance and covariance parameters $\{ {\bf \Sigma}_k, \sigma_k^2, \tau_k^2, \kappa_k \}$ need to be estimated for each $k = 1, \dots, K$. No estimates will be obtained for the local mean coefficient vector $\widetilde{\boldsymbol{\beta}}$, as all of the mean parameters will be estimated globally.

One final note regarding the estimation of the kernel matrices: the kernel matrix for mixture component location $k$ will be obtained through estimating the parameters of its spectral decomposition, namely $\lambda_1$, $\lambda_2$, and $\eta$, where 
\begin{equation} \label{anisomat}
{\bf \Sigma} = \left[ \begin{array}{cc} \cos(\eta) & - \sin(\eta)  \\ \sin(\eta) & \cos(\eta)  \end{array} \right] \left[ \begin{array}{cc} \lambda_1 & 0  \\ 0 & \lambda_2  \end{array} \right] \left[ \begin{array}{cc} \cos(\eta) & \sin(\eta)  \\ -\sin(\eta) & \cos(\eta)  \end{array} \right]
\end{equation}
(in the case that we have fixed $d=2$). Here, $\lambda_1$ and $\lambda_2$ are eigenvalues and represent squared ranges (such that $\lambda_1 >0$ and $\lambda_2>0$) and $\eta$ represents an angle of rotation, constrained to lie between $0$ and $\pi/2$ for identifiability purposes (\citealp{Katzfuss2013}).

The full model in Equation~\ref{model} can be fit after plugging REML estimates $\{ \widehat{\bf \Sigma}_k,  \widehat{\sigma}^2_k,  \widehat{\tau}^2_k,  \widehat{\kappa}_k : k = 1, \dots, K \}$ into the covariance function in Equation~\ref{NScov} using the discrete basis representation in Equation~\ref{spat_var_par} to calculate the likelihood for the observed data. Variance quantities that are not specified to be spatially-varying can then be estimated again using REML with the spatially-varying components considered fixed. For example, if for a particular model only ${\bf \Sigma}(\cdot)$ is allowed to vary spatially and the smoothness is fixed, it remains to estimate the overall nugget $\tau^2$ and variance $\sigma^2$. The restricted Gaussian likelihood for these parameters is then
\[
\mathcal{L}^R( \sigma^2, \tau^2 ; {\bf Z}, {\bf R} ) = -\frac{1}{2} \log \left| \sigma^2 {\bf R} + \tau^2 {\bf I}_n \right| - \frac{1}{2} \log |{\bf X}^\top (\sigma^2 {\bf R} + \tau^2 {\bf I}_n)^{-1} {\bf X} | - \frac{1}{2} {\bf Z}^\top {\bf P}{\bf Z},
\]
where ${\bf R}$ is the correlation matrix, i.e., the matrix calculated using Equation~\ref{NScov} without the $\sigma(\cdot)$ terms, and ${\bf P}$ is defined as in Equation~\ref{Pmat}. Once all of the covariance parameters have been estimated, the estimate of $\boldsymbol{\beta}$ can be calculated as in Equation~\ref{beta_GLS}.

Using this model requires both the number and placement of mixture component locations $\{ {\bf b}_k : k = 1, \dots, K\}$, selecting which of the spatial dependence parameters should be fixed or allowed to vary spatially, the tuning parameter for the weighting function $\lambda_w$, and the fitting radius $r$. Parameter estimates for this model are likely to be sensitive to the choice of $K$ and the placement of mixture component locations. Furthermore, \cite{TibshiraniHastie} discuss the importance of choosing $r$, which specifies the ``span size,'' suggesting that the model should be fit using a range of $r$ values, and use a global criterion such as the maximized overall likelihood, cross-validation, or Akaike's Information Criterion to choose the final model. This strategy could either be implemented on a trial-and-error basis or in an automated scheme. Of course, regardless of the number and locations of the mixture component centroids, the radius $r$ should be chosen such that a large enough number of data points are used to estimate a local stationary model. 

While different in both motivation and nature, the model outlined above is related to the local likelihood method described in \cite{Anderes_Stein}, which ties together locally stationary models to estimate a globally nonstationary model. The model in \cite{Anderes_Stein} involves optimizing a sum of weighted increments of local log-likelihoods, where the weights are estimated smoothly using a smoothing kernel. Alternatively, our approach estimates spatially-varying parameters locally using only a subset of the data, then fixing the global parameters according to Equation~\ref{spat_var_par}. Both of these approaches avoid the lack-of-smoothness issues innate to other similar segmentation approaches, such as \cite{fuentes2001} or the \textit{ad hoc} nonstationary kriging approach in \cite{PacScher}, which \cite{Anderes_Stein} call ``hard thresholding'' local likelihood estimates. Like \cite{Anderes_Stein}, our approach avoids the problem of non-smooth local parameter estimates implicit to hard thresholding methods by using the mixture component representation. 

We conclude this overview of our methodology with a note regarding the computational demands of fitting this model. Recall that Gaussian likelihood calculations such as Equation~\ref{full_lik} typically involve inverting and calculating the determinant of $n\times n$ matrices, requiring $\mathcal{O}(n^3)$ calculations (in this case, for each iteration of the optimization procedure). The local likelihood approach involves inverting a collection of $K$ $n_k \times n_k$ matrices, so that the model requires more like $\mathcal{O}(K\overline{n}^3)$ calculations, where $\overline{n} = K^{-1}\sum_k n_k$ is the average local sample size. This represents a significant reduction in both the required memory and CPU time, so long as $\overline{n} << n$. However, note that since the local models are fit independently of each other, if parallelization is utilized (see Section \ref{discussion}) then the computational time could be further reduced  to $\mathcal{O}(\overline{n}^3)$.

\section[Using the convoSPAT package for R]{Using the \pkg{convoSPAT} package for \proglang{R}} \label{sec5}

The \pkg{convoSPAT} package (version 1.1.1) is available CRAN, and can be installed as usual:

\begin{Sinput}
R> install.packages( "convoSPAT" )
R> library( "convoSPAT" )
\end{Sinput}

\noindent All of the data sets in Sections \ref{simulateddataeg} and \ref{USprecip} are included in the package. The \pkg{convoSPAT} package uses functionality from the \pkg{ellipse} (\citealp{R_ellipse}), \pkg{fields} (\citealp{R_fields}), \pkg{geoR} (\citealp{R_geoR}), \pkg{MASS} (\citealp{R_MASS}), \pkg{plotrix} (\citealp{R_plotrix}),  \pkg{sp} (\citealp{spBook}; \citealp{R_sp}), and \pkg{StatMatch} (\citealp{R_StatMatch}) packages for \proglang{R}. 

Two notes should be made before discussing the functionality of the package. First, while the methods described in Sections \ref{sec2}, \ref{NSGPM}, and \ref{CEI} are valid for spatial coordinates in a general $d$-dimensional Euclidean space, the following implementation only allows for two-dimensional coordinates, with $d=2$. Second, on a more technical note, the package is implemented using the S3-classes of \proglang{R}, and therefore contains package-specific \code{predict} and \code{plot} functionality.

\subsection{Nonstationary model fitting}

\sloppypar{
The primary components of the \pkg{convoSPAT} package are the \code{NSconvo_fit} and \code{predict.NSconvo} functions which fit the nonstationary model discussed in Section \ref{NSGPM} and provide predictions, respectively. 
}
The \code{NSconvo_fit} function takes the following arguments (with defaults as given):

\begin{verbatim}
NSconvo_fit( geodata = NULL, sp.SPDF = NULL,
  coords = geodata$coords, data = geodata$data,
  cov.model = "exponential", mean.model = data ~ 1, mc.locations = NULL, 
  N.mc = NULL, mc.kernels = NULL, fit.radius, lambda.w = NULL,
  ns.nugget = FALSE, ns.variance = FALSE, local.pars.LB = NULL, 
  local.pars.UB = NULL, global.pars.LB = NULL, global.pars.UB = NULL,  
  local.ini.pars = NULL, global.ini.pars = NULL )
\end{verbatim}

\noindent The spatial coordinates and response variable of interest may be input in several different ways: first, the function accepts a \code{geodata} object from the \pkg{geoR} package (\citealp{R_geoR}), using the \code{geodata} input; second, a \code{SpatialPointsDataFrame} object from the \pkg{sp} package (\citealp{spBook}; \citealp{R_sp}), using the \code{sp.SPDF} input; finally, the coordinates and data may be entered directly using the \code{coords} and \code{data} inputs. As a result, the package is able to handle a variety of geographic coordinate reference systems; note, however, that distances between points are calculated using a Mahalanobis distance (see Equation~\ref{Q}).

Two other required inputs are the number of mixture component locations (\code{N.mc}) and the \code{fit.radius} (previously denoted $r$). The user may specify a covariance model from the \pkg{geoR} (\citealp{R_geoR}) options \code{cauchy}, \code{matern}, \code{circular}, \code{cubic}, \code{gaussian}, \code{exponential}, \code{spherical}, or \code{wave}, as well as a mean model through the usual formula notation (a constant mean is the default). For most applications (and as an alternative to specifying \code{N.mc}), the user will want to specify the mixture component locations directly: the default is to create an evenly spaced grid over the spatial domain of interest, which may not be appropriate if the spatial domain is non-rectangular. The tuning parameter for the weighting function $\lambda_w$ is defined by \code{lambda.w}. The default for $\lambda_w$ is fixed to be the square of one-half of the minimum distance between mixture component locations, or $(0.5 \min \{ || {\bf b}_k - {\bf b}_{k'} || \} )^2$ (in order to ensure a default scaling appropriate for the resolution of the mixture component grid), but may also be specified by the user. The user may also specify if either the nugget variance or process variance is to be spatially-varying by setting either \code{ns.nugget = TRUE} or \code{ns.variance = TRUE} (or both). If the mixture component kernels themselves are pre-specified (e.g., based on expert opinion), these may also be passed into the function, which will greatly reduce computational time.

Note that if the data and coordinates are not specified as a \code{geodata} object, the \code{data} argument for this function can accommodate replicates. This might be of interest for applications similar to the ones in \cite{sampGutt}, in which the replicates represent repeated observations over time that have been temporally detrended. In this case, the model will assume a constant spatial dependence structure over replicates (time) as well as the same mean function over replicates (that is, the locations must be constant across replicates; furthermore, the regression coefficients will be constant across replicates).

The optimization method used within \code{optim} for this package is \code{"L-BFGS-B"}, which allows for the specification of upper and lower bounds for each parameter with respect to the optimization search. The upper and lower bounds may be passed to the function via \code{local.pars.LB}, \code{local.pars.UB}, \code{global.pars.LB}, and \code{global.pars.UB}. The local limits require vectors of length five, with bounds for the local parameters $\lambda_1, \lambda_2, \tau^2$, $\sigma^2$, and $\kappa$, while the global limits require vectors of length three, with bounds for the global parameters  $\tau^2$, $\sigma^2$, and $\kappa$. Default values for these limits are as follows: for both the global and local parameter estimation, the lower bounds for $\lambda_1, \lambda_2, \sigma^2, \tau^2$, and $\kappa$ are fixed at 1e-5; the upper bound for the smoothness $\kappa$ will be fixed to 30. The upper bounds for the variance and kernel parameters, on the other hand, will be specific to the application: for the nugget variance ($\tau^2$) and process variance ($\sigma^2$), the upper bound will be $4 \widehat{\sigma}^2_{OLS}$ (where $\widehat{\sigma}^2_{OLS}$ is the error variance estimate from a standard ordinary least squares procedure); the upper bound for $\lambda_1$ and $\lambda_2$ will be one-fourth of the maximum interpoint distance between observation locations in the data set. The bounds for $\eta$ are fixed at $0$ and $\pi/2$.

Given that many calls to \code{optim} are made within \code{NSconvo_fit}, the function prints a message to notify the user if \code{optim} returns any errors. In test runs of the package, the most common warning (non-fatal) message encountered is \code{"ABNORMAL_TERMINATION_IN_LINSRCH"}, which seems to have no negative impact on the results of the optimization.
 
 \sloppypar{
The final options in the \code{NSconvo_fit} function involve \code{local.ini.pars} and \code{global.ini.pars}, which specify the initial values used for the local and global calls of \code{optim}, respectively. As with the limits, \code{local.ini.pars} is a vector of length five, with initial values for the local parameters $\lambda_1, \lambda_2, \tau^2$, $\sigma^2$, and $\kappa$, while \code{global.ini.pars} is a vector of length three, with initial values for the global parameters  $\tau^2$, $\sigma^2$, and $\kappa$. The default for these inputs are as follows: $\lambda_{1,init} = \lambda_{2, init} = c/10$, where $c$ is the maximum interpoint distance between observation locations, $\tau^2_{init} = 0.1\widehat{\sigma}^2_{OLS}$, $\sigma^2_{init} = 0.9\widehat{\sigma}^2_{OLS}$, and $\kappa_{init} = 1$.
}

When the \code{NSconvo_fit} function is called, the status of the model fitting will be printed on the screen. As the function fits the locally stationary models for each mixture component location, the function will print the mixture component and number of observations that are currently being used for estimation. After the local models have been fit for each mixture component location, a printed message will notify the user that the variance parameters are being estimated globally (if applicable).

A function which may be helpful before running \code{NSconvo_fit} is the \code{mc_N} function, which returns the number of observations which will be used to fit each local model for a particular set of mixture component locations and fit radius. This function may be helpful when selecting the fit radius, as the user may want to know how many data points will be used to fit the local model for a number of different fit radii. The inputs to the function are

\begin{verbatim}
mc_N( coords, mc.locations, fit.radius )
\end{verbatim} 

\noindent where \code{coords} are the observation locations for the full data set, \code{mc.locations} are the mixture component locations, and \code{fit.radius} is the fitting radius. This function is also automatically run inside the \code{NSconvo_fit} wrapper, printing a warning message if any of the local sample sizes are less than 5. In this case, the user should refine the mixture component grid or expand the fit radius.

After the model fitting has completed, \code{NSconvo_fit} returns a \code{NSconvo} object which can be passed to the \code{summary.NSconvo} function to quickly summarize the fitted model. Among other elements, a \code{NSconvo} object includes:

\begin{itemize}
\item[] \code{mc.kernels}, which contains the estimated kernel matrices for the mixture component locations,
\item[] \code{mc.locations}, which contains the mixture component locations,
\item[] \code{MLEs.save}, which includes a data frame of the locally-estimated stationary model parameters for each mixture component location,
\item[] \code{kernel.ellipses}, which includes the estimated kernel ellipse for each location in \code{coords},
\item[] \code{beta.GLS}, the generalized least squares estimate of $\boldsymbol{\beta}$,
\item[] \code{beta.cov}, the estimated covariance matrix of $\widehat{\boldsymbol{\beta}}$,
\item[] \code{tausq.est}, the estimate of the nugget variance -- either a constant (if estimated globally) or a vector with the estimated nugget variance for each location in \code{coords},
\item[] \code{sigmasq.est}, the estimate of the process variance -- either a constant (if estimated globally) or a vector with the estimated process variance for each location in \code{coords},
\item[] \code{kappa.MLE}, the global estimate of the smoothness (for \code{cauchy} or \code{matern}),
\item[] \code{Cov.mat} and \code{Cov.mat.inv}, the estimated covariance matrix for the data and its inverse (respectively), and
\item[] \code{Xmat}, the design matrix for the mean model.
\end{itemize}

The \code{NSconvo} object can also be passed to the \code{predict.NSconvo} function, which calculates predictions $\widehat{\boldsymbol{\mu}}_{\bf Z^*|z}$ and prediction standard errors $\widehat{\bf \Sigma}_{\bf Z^*|z}$. The \code{predict.NSconvo} function takes the following arguments:

\begin{verbatim}
predict.NSconvo( object, pred.coords, pred.covariates = NULL, ... )
\end{verbatim}

\noindent The \code{object} is the output of \code{NSconvo_fit}, \code{pred.coords} is a matrix of the prediction locations of interest, \code{pred.covariates} is a matrix of covariates for the prediction locations (the intercept is added automatically), and \code{...} allows other options to be passed to the default \proglang{R} \code{predict} methods. Calculating the predictions when the dimension of \code{pred.coords} is large is computationally expensive, and a progress meter prints while the machine is working. The output from \code{predict.NSconvo} includes:

\begin{itemize}
\item[] \code{pred.means}, which contains the kriging predictor for each prediction location, and
\item[] \code{pred.SDs}, which contains the corresponding prediction standard error.
\end{itemize}

\subsection{Anisotropic model fitting} \label{aniso_fitting}

For the sake of comparison, the functions \code{Aniso_fit} and \code{predict.Aniso} are also provided, which fit the stationary (anisotropic) model with the covariance function given by Equation~\ref{ani_model} to the full dataset. Note: the following functions have been coded from scratch and represent reimplementations of methodology that already exists in other packages, such as \pkg{geoR} (\citealp{R_geoR}). In spite of being reimplementations, the functions are provided for convenience, as their syntax matches \code{NSconvo_fit} and can be quickly implemented in tandem with \code{NSconvo_fit}. However, also note that the version provided here has not been optimized nearly to the extent of the version in \pkg{geoR}, and therefore \code{Aniso_fit} will take significantly longer than a corresponding implementation of, say, \code{likfit} in \pkg{geoR}.

The \code{Aniso_fit} function takes the following arguments (with defaults as given):

\begin{verbatim}
Aniso_fit( geodata = NULL, sp.SPDF = NULL,
  coords = geodata$coords, data = geodata$data,
  cov.model = "exponential", mean.model = data ~ 1, local.pars.LB = NULL, 
  local.pars.UB = NULL, local.ini.pars = NULL )
\end{verbatim}

\noindent The inputs to this function are identical to the corresponding inputs to the nonstationary model fitting function, and the output is an \code{Aniso} object. Among other elements, an \code{Aniso} object includes:

{

\begin{itemize}
\item[] \code{MLEs.save}, which includes a data frame of the locally-estimated stationary model parameters for each mixture component location,
\item[] \code{beta.GLS}, the generalized least squares estimate of $\boldsymbol{\beta}$,
\item[] \code{beta.cov}, the estimated covariance matrix of $\widehat{\boldsymbol{\beta}}$,
\item[] \code{aniso.pars}, the global estimate of the anisotropy parameters $\lambda_1$, $\lambda_2$, and $\eta$, which define the anisotropy matrix in Equation~\ref{anisomat},
\item[] \code{aniso.mat}, which gives the global estimate of ${\bf \Sigma}$ from Equation~\ref{anisomat} in matrix form,
\item[] \code{tausq.est}, the global estimate of the nugget variance,
\item[] \code{sigmasq.est}, the global estimate of the process variance,
\item[] \code{kappa.MLE}, the global estimate of the smoothness (for \code{cauchy} or \code{matern}), and
\item[] \code{Cov.mat} and \code{Cov.mat.inv}, the estimated covariance matrix for the data and its inverse (respectively)
\end{itemize}
}

Once the anisotropic model has been fit and the output stored, the fitted model object can be passed to the \code{predict.Aniso} function, which (similar to the nonstationary predict function) calculates predictions $\widehat{\boldsymbol{\mu}}_{\bf Z^*|z}$ and prediction standard errors $\widehat{\bf \Sigma}_{\bf Z^*|z}$. The arguments are again identical to the nonstationary predict function: 

\begin{verbatim}
predict.Aniso( object, pred.coords, pred.covariates = NULL, ... )
\end{verbatim}

\noindent The \code{object} is the output of \code{Aniso_fit}, \code{pred.coords} is a matrix of the prediction locations of interest, and \code{pred.covariates} is a matrix of covariates for the prediction locations (the intercept is added automatically). Similar to the nonstationary predict function, the output from \code{predict.Aniso} includes:

\begin{itemize}
\item[] \code{pred.means}, which contains the kriging predictor for each prediction location, and
\item[] \code{pred.SDs}, which contains the corresponding prediction standard error.
\end{itemize}

\subsection{Evaluation criteria and plotting functions} \label{plotting}

This package includes a function to quickly calculate the evaluation criteria described in Section \ref{predic}, as well as functions to visualize various components of the nonstationary model.

First, the \code{evaluate_CV} function calculates the MSPE, from Equation~\ref{MSPE}, the pMSDR, from Equation~\ref{pMSDR}, and the CRPS, from Equation~\ref{CRPS}. The function inputs are simply

\begin{verbatim}
evaluate_CV( holdout.data, pred.mean, pred.SDs )
\end{verbatim}

\noindent where \code{holdout.data} is the held-out validation data and \code{pred.mean} and \code{pred.SDs} are output from one of the fit functions. Note that the user must perform the subsetting of the data. The output of \code{evaluate_CV} is simply the MSPE, pMSDR, and CRPS, averaged over all hold-out locations.

Next, plotting functions are provided to help visualize the output of either the stationary or nonstationary model. The first is \code{plot.NSconvo}:
\begin{verbatim}
plot.NSconvo( x, plot.ellipses = TRUE, fit.radius = NULL, 
  aniso.mat = NULL, true.mc = NULL, ref.loc = NULL, 
  all.pred.locs = NULL, grid = TRUE, true.col = 1, 
  aniso.col = 4, ns.col = 2, plot.mc.locs = TRUE, ... )
\end{verbatim}
which plots either the estimated anisotropy ellipses (\code{plot.ellipses = TRUE}) or the estimated correlation (\code{plot.ellipses = FALSE}). The \code{x} is a \code{NSconvo} object; additional options can be added to a plot of the estimated anisotropy ellipses by specifying the \code{fit.radius} that was used to fit the model, the ellipse for the stationary model \code{aniso.mat} estimated in \code{Aniso_fit}, and the true mixture component ellipses (if known). To plot the estimated correlation, a reference location (\code{ref.loc}) must be specified, as well as all of the prediction locations of interest (\code{all.pred.locs}) and whether or not the predictions lie on a rectangular grid. The other options correspond to color values for the true mixture component ellipses (\code{true.col}), the anisotropy ellipse (\code{aniso.col}), and the estimated mixture component ellipses (\code{ns.col}); \code{plot.mc.locs} indicates whether or not the mixture component locations should be plotted. Note that the plotted ellipse is the 0.5 probability ellipse for a bivariate Gaussian random variable with covariance equal to the kernel matrix.

Similarly, the \code{plot.Aniso} function is provided to plot the estimated correlations from the stationary model. The function is
\begin{verbatim}
plot.Aniso( x, ref.loc = NULL, all.pred.locs = NULL, grid = TRUE, ... )
\end{verbatim}
The inputs are identical to the \code{plot.NSconvo} function, except that the \code{object} must be of the \code{Aniso} class.

\subsection{Other functions}

Two additional functions are also provided to simulate data from the nonstationary model discussed in Section \ref{NSGPM}, and are used to create the simulated data set in Section \ref{simulateddataeg}. First is \code{f_mc_kernels}, which calculates the true mixture component kernel matrices through a generalized linear model for each component of the kernel matrices' spectral decomposition as given in Equation~\ref{anisomat}. The function, with default settings, is

\begin{verbatim}
f_mc_kernels( y.min = 0, y.max = 5, x.min = 0, x.max = 5, N.mc = 3^2, 
  lam1.coef = c( -1.3, 0.5, -0.6 ), lam2.coef = c( -1.4, -0.1, 0.2 ), 
  logit.eta.coef = c( 0, -0.15, 0.15 ) )
\end{verbatim}

\noindent The inputs \code{y.mon}, \code{y.max}, \code{x.min}, and \code{x.max} define a rectangular spatial domain, \code{N.mc} is the number of mixture component locations, and \code{lam1.coef}, \code{lam2.coef}, and \code{logit.eta.coef} define regression coefficients for the spatially-varying parameters $\lambda_1$, $\lambda_2$, and $\eta$. For example, taking \code{lam1.coef} $\equiv \boldsymbol{\beta}_{\lambda_1} = (\beta_0^{\lambda_1}, \beta_1^{\lambda_1}, \beta_2^{\lambda_1})^\top$, the first eigenvector for an arbitrary location ${\bf s} = (s_1, s_2)$ is
\[
\lambda_1({\bf s}) = \exp \{ \beta_0^{\lambda_1} +  \beta_1^{\lambda_1} s_1 +  \beta_2^{\lambda_1}s_2 \}.
\]
The other eigenvalue is calculated similarly. The angle of rotation, on the other hand, is calculated through the scaled inverse logit transformation
\[
\eta({\bf s}) = \frac{\pi}{2} \cdot \text{logit}^{-1} \big( \beta_0^{\eta} +  \beta_1^{\eta} s_1 +  \beta_2^{\eta}s_2  \big),
\]
where \code{logit.eta.coef} $\equiv \boldsymbol{\beta}_{\eta} = (\beta_0^{\eta}, \beta_1^{\eta}, \beta_2^{\eta})^\top$. The default coefficients are those used to generate the simulated data set in Section \ref{simulateddataeg}, and were obtained by trial and error. The output of this function includes

\begin{itemize}
\item[] \code{mc.locations}, which contains the mixture component locations, and
\item[] \code{mc.kernels}, which contains the mixture component kernels for each mixture component location.
\end{itemize}

The true mixture component kernel matrices generated in \code{f_mc_kernels} can be used to simulate a data set using the function

\begin{verbatim}
NSconvo_sim( grid = TRUE, y.min = 0, y.max = 5, x.min = 0, x.max = 5, 
  N.obs = 20^2, sim.locations = NULL, mc.kernels.obj = NULL,
  mc.kernels = NULL, mc.locations = NULL, tausq = 0.1, sigmasq = 1, 
  beta.coefs = 4, kappa = NULL, covariates = rep( 1, N.obs ), 
  cov.model = "exponential" )
\end{verbatim}

\noindent In this function, \code{grid} is a logical input specifying if the simulated data should lie on a grid (\code{TRUE}) or not (\code{FALSE}), \code{y.min}, \code{y.max}, \code{x.min}, and \code{x.max} define the rectangular spatial domain, \code{N.obs} specifies the number of observed locations, \code{mc.kernels.obj} is an object from \code{f_mc_kernels}, \code{tausq}, \code{sigmasq}, \code{beta.coefs}, and \code{kappa} specify the true parameter values, \code{covariates} specifies the design matrix for the mean function, and \code{cov.model} specifies the covariance model. The output of this function is

\begin{itemize}
\item[] \code{sim.locations}, which contains the simulated data locations, 
\item[] \code{mc.locations}, which contains the mixture component locations, 
\item[] \code{mc.kernels}, which contains the mixture component kernels for each mixture component location,
\item[] \code{kernel.ellipses}, which contains the kernel matrices for each simulated data location,
\item[] \code{Cov.mat}, which contains the true covariance matrix of the simulated data, and
\item[] \code{sim.data}, which contains the simulated data.
\end{itemize}

\section{Example 1: Simulated data} \label{simulateddataeg}

\begin{figure}[!t]
    \centering
    \begin{minipage}{.5\textwidth}
        \centering
	\includegraphics[trim={0 40 0 20mm}, clip, width=0.97\textwidth]{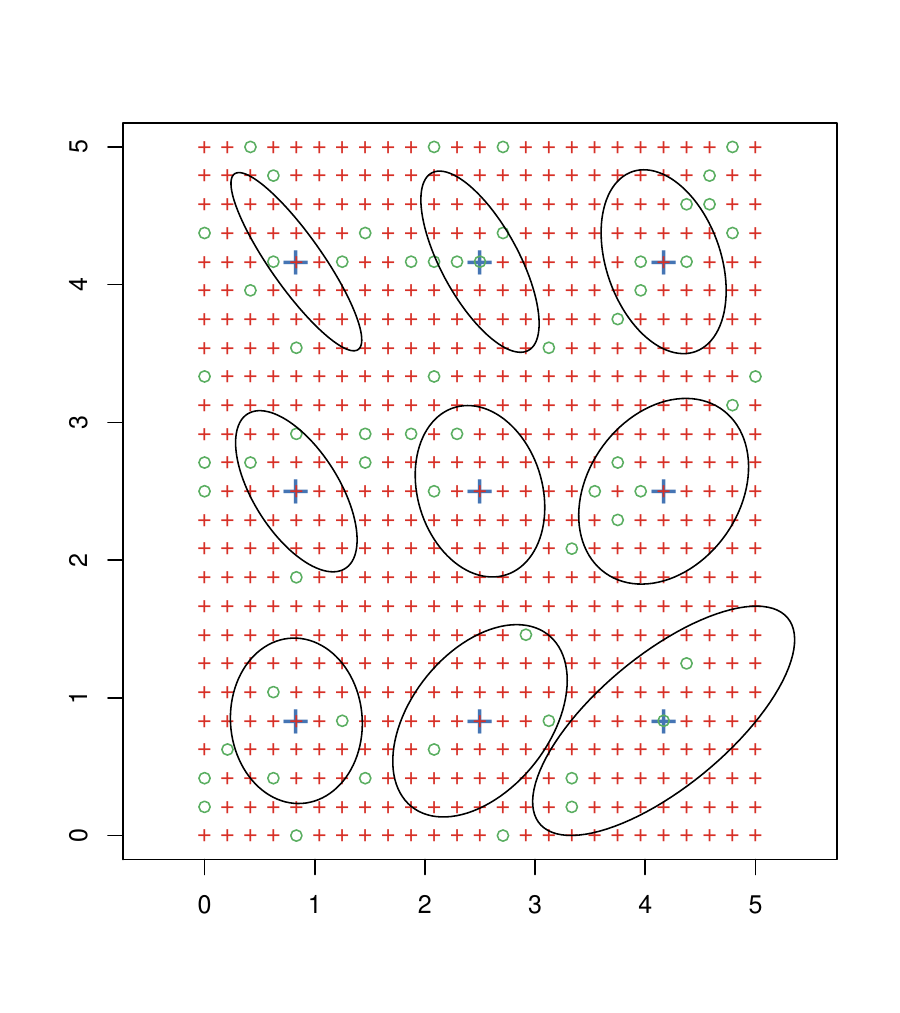}
    \end{minipage}%
    \begin{minipage}{0.5\textwidth}
        \centering
	\includegraphics[trim={0 40 0 16.75mm}, clip, width=\textwidth]{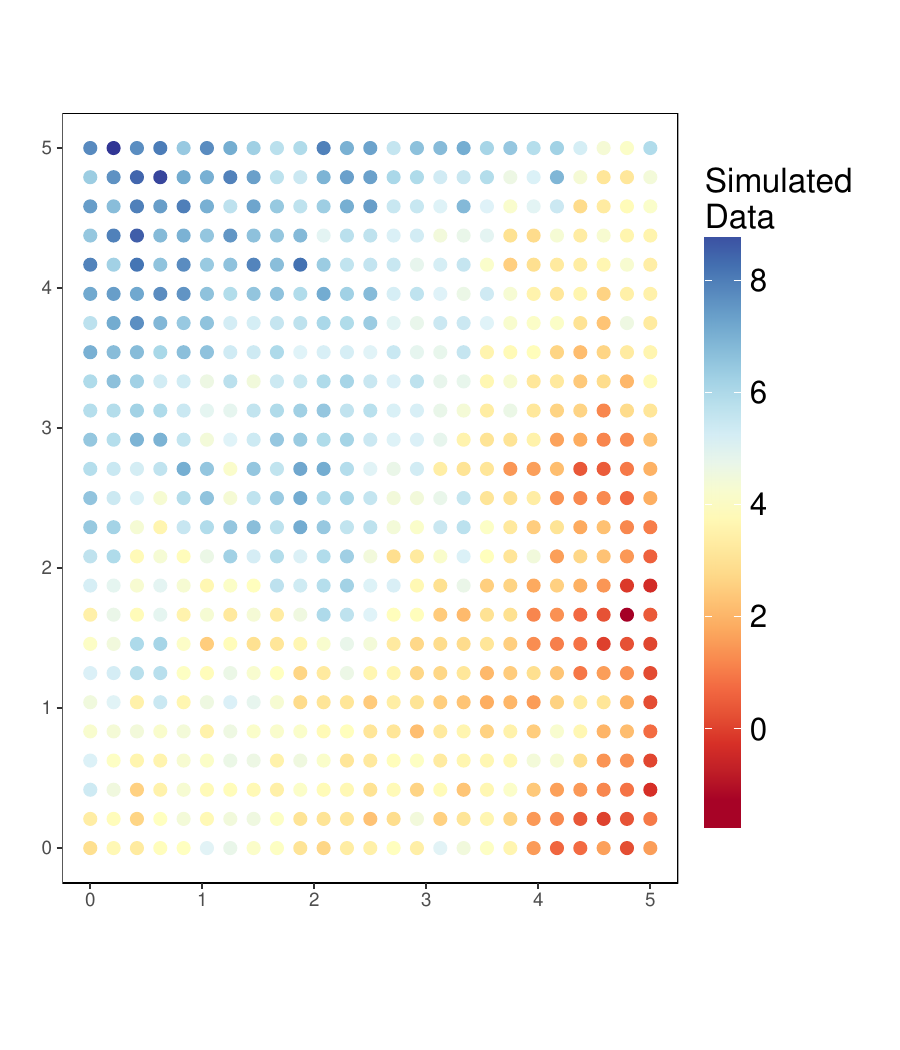}
    \end{minipage}
\caption{Left: true mixture component ellipses with observation locations (red) and validation locations (green). Right: simulated data.} 
\label{sim_data}
\end{figure}

As a simple illustration, the nonstationary model will be fit to an artificial data set simulated from the model. The data lie on a 25 $\times$ 25 grid (so that $n = 25^2 = 625$), and there are $K=9$ mixture component locations with corresponding mixture component ellipses as given in Figure \ref{sim_data}. Only the kernel matrices are allowed to vary spatially, an exponential correlation structure is used, and the mean structure contains the main effects of both coordinates. The true parameter values are $\tau^2 = 0.1$, $\sigma^2 = 1$, $\beta_0 = 4$, $\beta_{1} = -0.5$ ($x$-coordinate coefficient), $\beta_{2} = 0.5$ ($y$-coordinate coefficient), and $\lambda_w = 2$. A total of $m=60$ of the simulated data points are used as a validation sample. Figure \ref{sim_data} also provides the simulated data along with the validation locations.

The \code{simdata} object includes \code{sim.locations}, the simulated data locations; \code{mc.locations}, the mixture component locations; \code{mc.kernels}, the true mixture component kernel matrices; \code{sim.data}, the simulated data; and \code{holdout.index}, a vector of the 60 randomly sampled validation location indices. Note that there are actually ten independent and identically distributed replicates of the data contained in the ten columns of \code{sim.data}; in what follows the first column of data will be used. 

Figure \ref{sim_data} can be created with 

\begin{Sinput}
R> plot( simdata$mc.locations, pch = "+", asp = 1.25, xlab = "",
+    ylab = "", xlim=c( -0.5, 5.5 ), cex = 2, col = "#4575b4", main = " " )
R> points( simdata$sim.locations[ -simdata$holdout.index, ], col = "#d73027", 
+    pch = "+" )
R> points( simdata$sim.locations[ simdata$holdout.index, ], col = "#5aae61" )
R> for( i in 1:dim( simdata$mc.locations )[1] ){
+    lines( ellipse( simdata$mc.kernels[ , , i ],
+    centre = simdata$mc.locations[ i, ], level = 0.5 ) )
R> ggplot( data.frame( simdata = simdata$sim.data[ , 1 ],
+    xcoord = simdata$sim.locations[ , 1 ],
+    ycoord = simdata$sim.locations[ , 2 ] ), 
+    aes( x = xcoord, y = ycoord, color = simdata ) ) +
+    coord_fixed( ratio = 1.25 ) + geom_point( size = 2.5 ) + 
+    xlab( "" ) + ylab( "" ) + name = "Simulated \nData") +
+    scale_color_gradientn( colours = brewer.pal( 11, "RdYlBu" ),
+    theme( panel.background = element_rect( fill = "white" ),
+    panel.border = element_rect( colour = "black", fill = NA ),
+    panel.grid = element_blank(), legend.key.height = unit( 2, "cm" ),
+    legend.title = element_text( size = 16 ),
+    legend.text = element_text( size = 15 ) )
\end{Sinput}

\noindent (Note: this code requires the \pkg{ggplot2} (\citealp{R_ggplot2}), \pkg{RColorBrewer} (\citealp{R_RColorBrewer}), and \pkg{ellipse} (\citealp{R_ellipse}) packages in \proglang{R}.) 

\subsection{Selection of fixed components in the model}

In order to fit the nonstationary model, the user must specify three components: (1) the number and placement of mixture component locations, (2) the fitting radius $r$, and (3) the tuning parameter $\lambda_w$. For this simulated data set, the true number and placement of the mixture component locations as well as the tuning parameter are known, and these true values will be used. However, even for the simulated data, an optimal fit radius is \textit{not} known. When the tuning parameter is unknown the package provides a default value, but this may or may not be the optimal choice. 

Given that the nonstationary model can be fit relatively quickly for given values of $r$, a recommended strategy is to fit the model many times for a variety of different values  (same for the mixture component grid and $\lambda_w$, when these are unknown) and choose the final values based on which combination yields the ``best'' results. 
Of course, there is a trade-off between choices of the mixture component grid and fit radius: for a particular grid, the radius should be chosen such that a reasonable number of data points are used to fit each locally stationary model. It is here that the \code{mc_N} function may be helpful, because the user can quickly determine how many locations fall within the fitting radius of each grid point. For example, using the simulated data:
\begin{Sinput}
R> mc_N( coords = simdata$sim.locations[ -simdata$holdout.index, ], 
+    mc.locations = simdata$mc.locations, fit.radius = 2 )
[1] 164 212 165 209 269 212 153 204 157
\end{Sinput}
That is, using the true mixture component grid and a fit radius of $r=2$, the local models will be fit using anywhere from 157 to 269 data points.

For the simulated data in this application, the ``best'' fit was chosen based on the quality of the anisotropy parameter estimates relative to the true values; in other cases not involving simulated data, ``best'' might be defined in terms of the cross-validation criteria (see, e.g., Section \ref{USprecip}). The values used for these three components in the final fit (Section \ref{final_sim}) were the true mixture component grid (with $K=9$ evenly spaced locations over the domain), a fit radius of $r=2.3$ units, and the true tuning parameter value of $\lambda_w = 2$.

\subsection{Final model fitting} \label{final_sim}

Using the final choices of the fixed model components, the nonstationary model can be fit to the non-hold-out data (and results summarized) by

\begin{Sinput}
R> NSfit.model <- NSconvo_fit(
+    coords = simdata$sim.locations[ -simdata$holdout.index, ],
+    data = simdata$sim.data[ -simdata$holdout.index, 1 ],
+    cov.model = "exponential", fit.radius = 2.3, lambda.w = 2,
+    mc.locations = simdata$mc.locations,
+    mean.model = simdata$sim.data[ -simdata$holdout.index, 1 ]
+    ~ simdata$sim.locations[ -simdata$holdout.index, 1 ]
+    + simdata$sim.locations[ -simdata$holdout.index, 2 ] )
R> summary( NSfit.model )
\end{Sinput}

\noindent Similarly, the anisotropic model can be fit to the non-hold-out data (and results summarized) by

\begin{Sinput}
R> anisofit.model <- Aniso_fit(
+    coords = simdata$sim.locations[ -simdata$holdout.index, ],
+    data = simdata$sim.data[ -simdata$holdout.index, 1 ],
+    cov.model = "exponential",
+    mean.model = simdata$sim.data[ -simdata$holdout.index, 1 ]
+    ~ simdata$sim.locations[ -simdata$holdout.index, 1 ]
+    + simdata$sim.locations[ -simdata$holdout.index, 2 ] )
R> summary( anisofit.model )
\end{Sinput}

\begin{table}[t]
\begin{center}
\begin{tabular}{|c|c|c|c|}
\hline  						
 & \textbf{True value} & \textbf{Stationary model} & \textbf{Nonstationary model}  \\
\hline \hline
$\beta_0$ (intercept)& 	4 	& 4.039  	& 3.905  \\ \hline
$\beta_1$ ($x$-coordinate) & 	-0.5 	&  -0.695	& -0.678  \\ \hline
$\beta_2$ ($y$-coordinate) & 	0.5 	&   0.741	& 0.770 \\ \hline
$\tau^2$ (nugget) & 	0.1 	&   0.086	&  0.107 \\ \hline
$\sigma^2$ (variance) & 	1 	&  1.038	&  0.958 \\ \hline \hline
CRPS & -- &   -0.424 & -0.415 \\ \hline
MSPE & -- & 0.546  & 0.524 \\ \hline
\end{tabular}
\caption{ Parameter estimates from the simulated data, comparing the stationary and nonstationary models.}
\label{sim_estimates}
\end{center}
\end{table}

Predictions for the validation locations under each model can be calculated by 

\begin{Sinput}
R> pred.NS <- predict( NSfit.model, 
+    pred.coords = simdata$sim.locations[ simdata$holdout.index, ],
+    pred.covariates = simdata$sim.locations[ simdata$holdout.index, ] )
R> pred.S <- predict( anisofit.model,
+    pred.coords = simdata$sim.locations[ simdata$holdout.index, ],
+    pred.covariates = simdata$sim.locations[ simdata$holdout.index, ] )

\end{Sinput}

\noindent (note: here, \code{predict} is an alias for \code{predict.NSconvo} and \code{predict.Aniso} for objects of these classes; similar syntax will be used throughout) after which the evaluation criteria can be calculated by

\begin{Sinput}
R> evaluate_CV( holdout.data = simdata$sim.data[ simdata$holdout.index ], 
+    pred.mean = pred.NS$pred.means, pred.SDs = pred.NS$pred.SDs )
R> evaluate_CV( holdout.data = simdata$sim.data[ simdata$holdout.index ], 
+    pred.mean = pred.S$pred.means, pred.SDs = pred.S$pred.SDs )
\end{Sinput}

\noindent A summary of the parameter estimates and cross-validation results (MSPE and CRPS only) from each fitted model is provided in Table \ref{sim_estimates}. 

Calculating predictions on a finer resolution can be done as follows:

\begin{Sinput}
R> grid.x <- seq( from = 0, to = 5, by = 0.05 )
R> grid.y <- seq( from = 0, to = 5, by = 0.05 )
R> grid.locations <- expand.grid( grid.x, grid.y )
R> pred.locs <- matrix( c( grid.locations[ , 1 ], grid.locations[ , 2 ] ), 
+    ncol = 2, byrow = F )
R> pred.NS.all <- predict( NSfit.model, pred.coords = pred.locs,
+    pred.covariates = pred.locs )
R> pred.S.all <- predict( anisofit.model, pred.coords = pred.locs,
+    pred.covariates = pred.locs )
\end{Sinput}

\begin{figure}[!t]
\begin{center}
\includegraphics[width=0.9\textwidth]{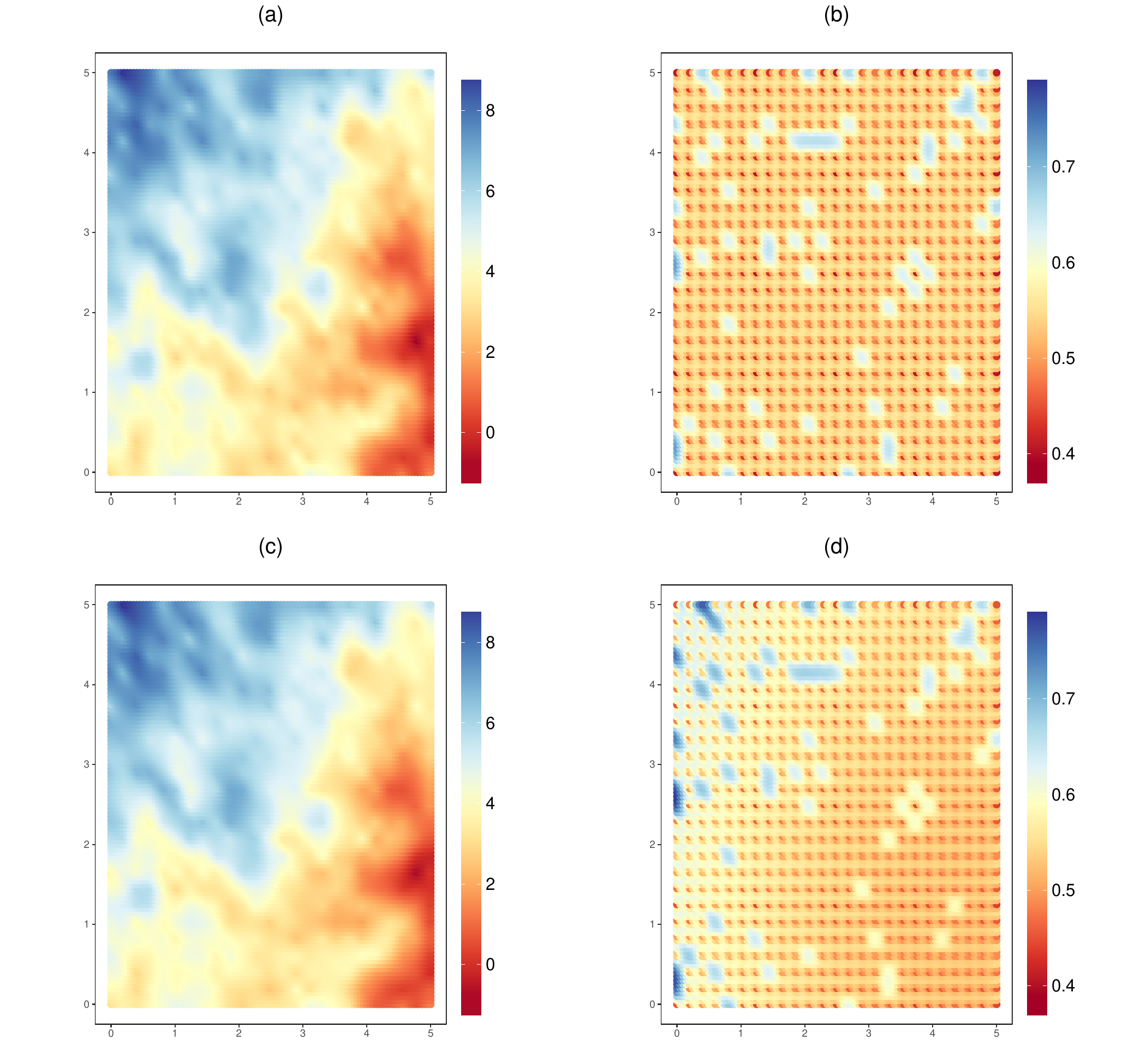}
\caption{Predictions and prediction errors from the stationary model (a. and b.) and the nonstationary model (c. and d.).} 
\label{stat_fits}
\end{center}
\end{figure}

\noindent Interpolated maps for the nonstationary model can be created with \pkg{ggplot2} (\citealp{R_ggplot2}):

\begin{Sinput}
R> ggplot( data.frame( preds = pred.NS.all$pred.means,
+    xcoord = pred.locs[,1],  ycoord = pred.locs[,2] ), 
+    aes( x = xcoord, y = ycoord, color = preds ) ) +
+    coord_fixed( ratio = 1.25 ) + geom_point( size = 2.5 ) + 
+    scale_color_gradientn( colours = brewer.pal( 11, "RdYlBu" ),
+    name = "", limits = c( -1.15, 8.7 ) ) +
+    theme( panel.background = element_rect( fill = "white" ),
+    panel.border = element_rect( colour = "black", fill = NA ),
+    panel.grid = element_blank() )
R> ggplot( data.frame( preds = pred.NS.all$pred.SDs,
+    xcoord = pred.locs[,1], ycoord = pred.locs[,2] ), 
+    aes( x = xcoord, y = ycoord, color = preds ) ) +
+    coord_fixed( ratio = 1.25 ) + geom_point( size = 2.5 ) + 
+    scale_color_gradientn( colours = brewer.pal( 11, "RdYlBu" ),
+    name = "", limits = c( 0.38, 0.78 ) ) +
+    theme( panel.background = element_rect( fill = "white" ),
+    panel.border = element_rect( colour = "black", fill = NA ),
+    panel.grid = element_blank() )
\end{Sinput}

(similarly for the stationary predictions and standard errors); these plots are provided in Figure \ref{stat_fits}. To visualize the locally-estimated anisotropy, the mixture component kernel matrices can be plotted along with the stationary anisotropy ellipse using

\begin{Sinput}
R> plot( NSfit.model, fit.radius = 2.3, xlab = "", ylab = "",
+    aniso.mat = anisofit.model$aniso.mat, asp = 1.25, 
+    true.mc = simdata$mc.kernels, true.col = 1,
+    aniso.col = 4, ns.col = 2, xlim = c( -1, 6 ), ylim = c( -1, 6 ) )
\end{Sinput}

\begin{figure}[!t]
\begin{center}
\includegraphics[trim={58.5 72.5 20 20mm}, clip, width=0.85\textwidth]{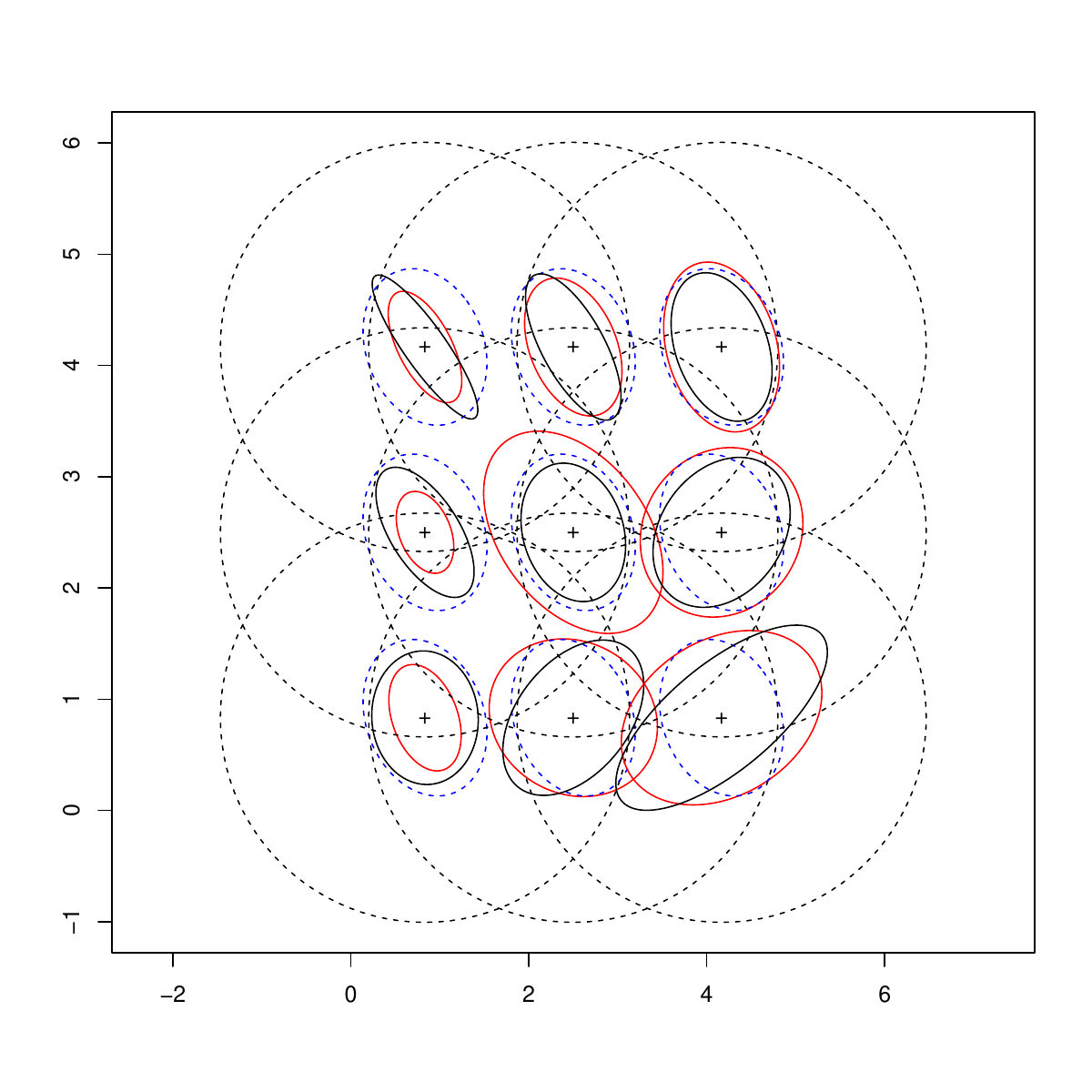}
\caption{True mixture component ellipses (solid black) with fit radius (dashed gray), nonstationary ellipses (solid red), and the stationary ellipse (dashed blue).} 
\label{est_ellip}
\end{center}
\end{figure}

\noindent (Note: as with \code{predict}, \code{plot} is an alias for \code{plot.NSconvo} for an object of class \code{NSconvo}; similar syntax will be used throughout.) This plot is shown in Figure \ref{est_ellip}; since the true anisotropy ellipses are known, these are also plotted. Note that the estimated nonstationary ellipses (solid red) compare quite favorably with the true ellipses (black); furthermore, the stationary ellipse (dashed blue) appears to be approximately an ``average'' of the spatially-varying ellipses.

\begin{figure}[!t]
\begin{center}
\includegraphics[trim={25 20 0 15mm}, clip, width=\textwidth]{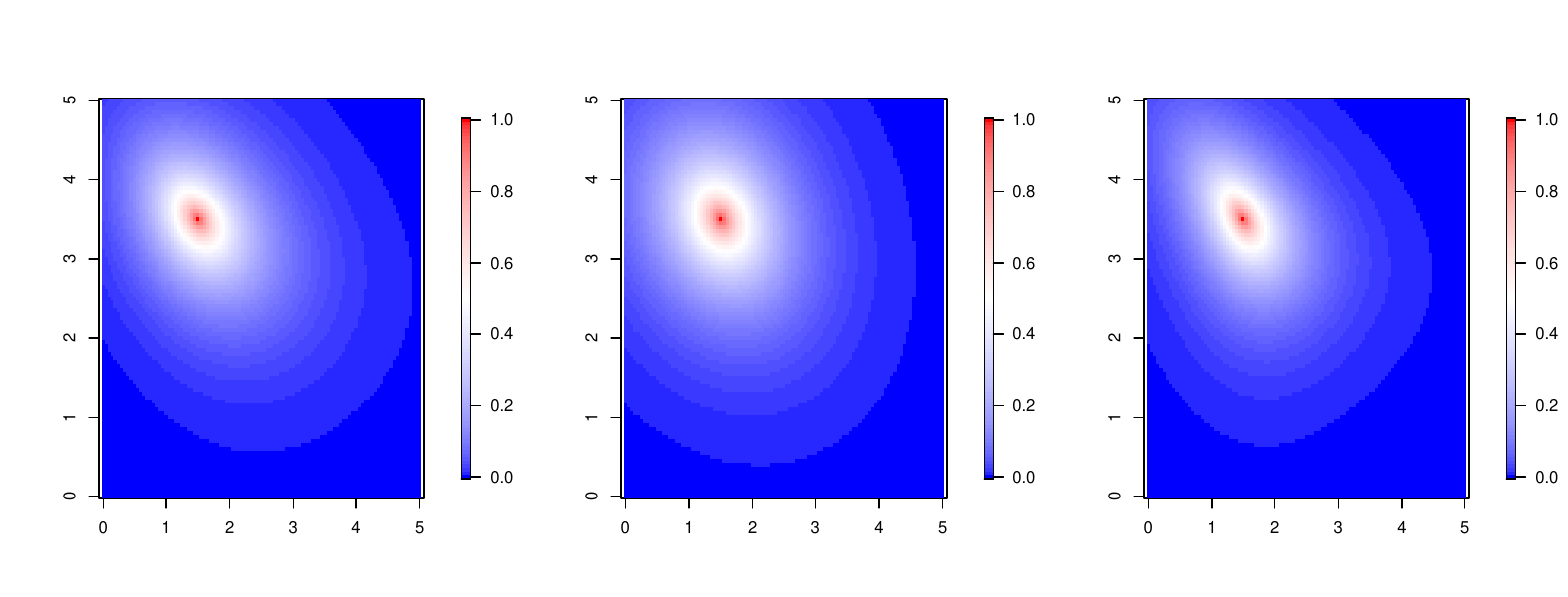}
\caption{Estimated correlations for a reference point, showing the nonstationary (left) and stationary (center) models, as well as the true correlation (right).} 
\label{corrplots}
\end{center}
\end{figure}

As an additional visualization of the estimated nonstationarity, estimated correlation plots for a particular reference point can be obtained by

\begin{Sinput}
R> plot( NSfit.model, plot.ellipses = FALSE, asp = 1.25,  
+    ref.loc = c( 1.5, 3.5 ), all.pred.locs = pred.locs, 
+    col = diverge_hsv( 100 ) )
R> plot( anisofit.model, asp = 1.25, ref.loc = c( 1.5, 3.5 ), 
+    all.pred.locs = pred.locs, col = diverge_hsv( 100 ) )
\end{Sinput}

\noindent and are given in Figure \ref{corrplots} for the nonstationary and stationary models, as well as the true correlation for comparison. Note that the stationary model estimates an elliptical correlation pattern, while the nonstationary model captures the non-elliptical nature of the true correlation pattern. 

Finally, note that the nonstationary model outperforms the stationary model in terms of both CRPS and MSPE (see Table \ref{sim_estimates}).

\begin{figure}[!t]
\begin{center}
\includegraphics[trim={5 5 5 5mm}, clip, width=\textwidth]{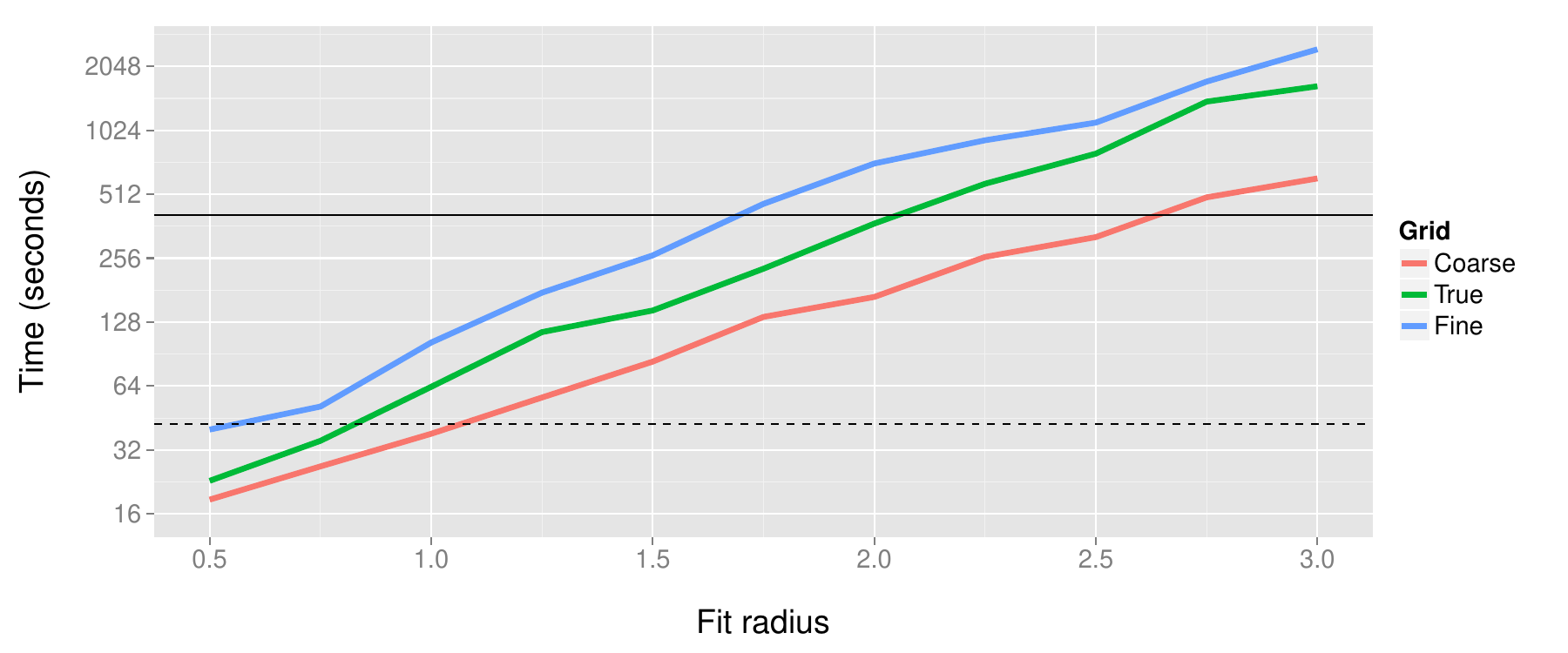}
\caption{Computational time (in seconds) for fitting the nonstationary model to the simulated data (of size $n = 565$) across three different mixture component grids (of size $K = 4, 9, 16$, respectively) and a range of fit radii. The solid and dashed black lines represents the computational time needed to fit the stationary model using, respectively, the \pkg{convoSPAT} and \pkg{geoR} packages. Computational times given are for a Dual Quad Core Xeon 2.66GHz machine with 32GB RAM.}
\label{comp_times}
\end{center}
\end{figure}

\subsection{Computational considerations}

Selection of the size and placement of the mixture component grid and the fit radius have a major impact on the computational time. As a demonstration of this relationship, consider the computational times for a variety of choices for the mixture component grid and fit radius, summarized in Figure \ref{comp_times}. Three mixture component grids were chosen: ``coarse'' (with $K=4$), ``true'' (with $K=9$), and ``fine'' (with $K=16$). For each of these grids, the nonstationary model (with spatially-varying nugget and variance) was fit using fit radii ranging between $r=0.5$ and $r=3$. As a reference point, the average number of data points used to fit the locally stationary models increases linearly from around 20 for $r=0.5$ to 350 for $r=3$ (consistent across the different mixture component grids).

The linear trends on the log scale in Figure \ref{comp_times} are expected, as calculations for Gaussian process models are known to increase with the square of the sample size. The (approximately) vertical shifts present in moving from coarse to true to fine are also expected, as the finer grids represent fitting more local models with (approximately) the same local sample sizes. 

For comparison, the time required to fit the stationary model using \code{Aniso_fit} is also shown on Figure \ref{comp_times} as a solid black line. However, as mentioned in Section \ref{aniso_fitting}, if the user is interested in fitting a stationary model to a spatial data set, a much better choice is \code{likfit} in \pkg{geoR} (\citealp{R_geoR}). The corresponding stationary (anisotropic) model in \pkg{geoR} can be fit by
\begin{Sinput}
R> geoR.fit <- likfit( data = simdata$sim.data[ -simdata$holdout.index, 1 ],
+    coords = simdata$sim.locations[ -simdata$holdout.index, ],
+    cov.model = "exponential", ini.cov.pars = c( 1, 1 ), trend = "1st",
+    fix.psiA = FALSE, fix.psiR = FALSE, lik.method = "REML" )
\end{Sinput}
The computational time for \code{likfit} is also shown on Figure \ref{comp_times} as a dashed black line.


\begin{figure}[!t]
\begin{center}
\includegraphics[trim={5 75 0 30mm}, clip, width=0.75\textwidth]{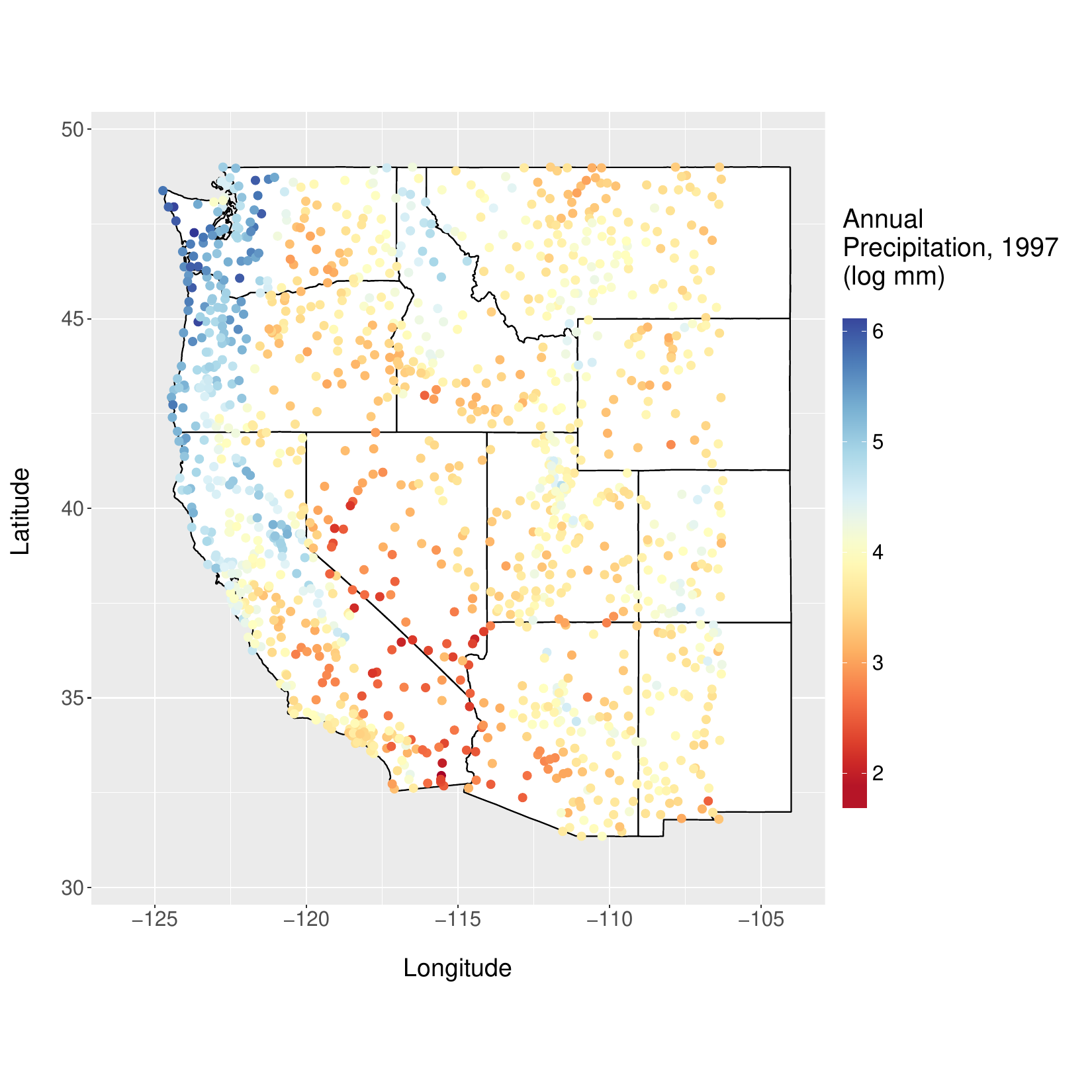}
\caption{Annual precipitation data for 1997, in log mm.}
\label{Precip_data}
\end{center}
\end{figure}

\section{Example 2: Annual precipitation} \label{USprecip}

The nonstationary model proposed in Sections \ref{NSGPM} and \ref{CEI} is also applied to a moderately large, real data set, consisting of the total annual precipitation in the western United States for 1997. The data is available online from the National Center for Atmospheric Research (\url{http://www.image.ucar.edu/GSP/Data/US.monthly.met}) as part of a larger data set that includes measurements for the entire United States. For the purposes of this analysis, a subset of the data that includes the western United States was chosen (see Figure \ref{Precip_data}) because precipitation is smoother and more densely observed over the central and eastern United States. This subset is included in the package as an \code{RData} (named \code{USprecip97}) file and consists of $1270$ observations. For illustration purposes, we work with non-projected longitude and latitude, but note that due to the size of the study area this may be inappropriate.

\subsection{Spatial model summaries} \label{modelSumm}

A total of five spatial models were fit to this dataset, the details of which are summarized in Table \ref{model_summary}: the stationary model and four nonstationary models. All models were assigned the same mean structure, which included the main effects of longitude and latitude as well as an intercept. Additionally, all models were assigned the same underlying correlation structure, chosen to be exponential (i.e., the Mat\'ern class with fixed $\kappa = 0.5$). Twenty percent of the observations ($m=254$) were held out as a validation data set in order to evaluate each model, leaving $n= 1016$ observations as a training data set. The stationary model was fit using the \code{Aniso_fit} function and the nonstationary models were fit using the \code{NSconvo_fit} function; the four nonstationary models represent all combinations of the \code{ns.nugget} and \code{ns.variance} options.

\begin{table}[!t]
\begin{center}
\begin{tabular}{|c|c|}
\hline  						
\textbf{Label} & \textbf{Covariance model details} \\
\hline
Stationary & Anisotropic, constant nugget and variance 		 \\ \hline
NS1 & SV anisotropy, constant nugget and variance 	 \\ \hline
NS2 & SV anisotropy and variance, constant nugget  \\ \hline
NS3 & SV anisotropy and nugget, constant variance  \\ \hline
NS4 & SV anisotropy, variance, and nugget  		 \\ \hline
\end{tabular}
\caption{ A brief summary of the different models fit to the precipitation data. Note: ``SV'' indicates ``spatially-varying.''}
\label{model_summary}
\end{center}
\end{table}

In addition to selecting which of the nugget variance and/or process variance should be spatially-varying, recall that using this model requires specifying (1) a mixture component grid, (2) the tuning parameter for the weight function, and (3) the fitting radius $r$. For each of the nonstationary models, the model was actually fit multiple times using two different mixture component grids (``coarse'', with $K=15$, and ``fine'', with $K=22$), four different values of the tuning parameter ($\lambda_w = 1.00, 2.67, 4.33, 6.00$), and six different fit radii (for the coarse grid, $r = 2.5, 3.2, 3.9, 4.6, 5.3, 6.0$; for the fine grid, $r=2.75, 3.20, 3.65, 4.10, 4.55, 5.00$).

Of the $2\times 4 \times 6 = 48$ total models fit for each of the four nonstationary models, the best model was selected based on maximizing the CRPS criteria (for the test data using the log precipitation); the best models are summarized in Table \ref{realdata_summary}. For each model, the coarse grid with a fit radius of $r=3.9$ performed best, and the largest tuning parameter was preferred for all but model NS1 (however, the CRPS for NS1 with $\lambda_w=6$ was -0.13714, which is only slightly smaller than the CRPS for $\lambda_w=4.33$, which was -0.13709). Table \ref{realdata_summary} also provides the computational run time for each model. Parameter estimates for each of the best models are summarized in Table \ref{param_estimates}. 

\begin{table}[!t]
\begin{center}
\begin{tabular}{|c||c|c|c|c|c|c|}
\hline  						
\textbf{Model} & \textbf{Grid size} & ${\lambda}_w$ & $r$ & \textbf{CRPS} & \textbf{MSPE} & \textbf{Comp. time} \\
\hline \hline
Stationary      & --	& --  & --   & -0.1424		&  0.0666 &   85.94 min. \\ \hline \hline
NS1 & 15 (coarse grid) 	& 4.33  & 3.9  &  -0.1371	&  0.0610	& 17.02 min. \\ \hline
NS2 & 15 (coarse grid) & 6.0  & 3.9  &  -0.1387	&  0.0619 & 16.31 min. \\ \hline
NS3 & 15 (coarse grid) &  6.0 & 3.9  &  -0.1377	&  0.0619 & 16.55 min. \\ \hline
NS4 & 15 (coarse grid) &  6.0 & 3.9  &  -0.1386	&  0.0623 & 9.49 min. \\ \hline
\end{tabular}
\caption{ Model details and evaluation for the best model of each type fit to the precipitation data, selected based on maximizing CRPS. The computational time given is for a Dual Quad Core Xeon 2.66GHz machine with 32GB RAM.}
\label{realdata_summary}
\end{center}
\end{table}

\begin{table}[!t]
\begin{center}
\begin{tabular}{|c||c||c|c|c|c|}
\hline  						
\textbf{Parameter} & \textbf{S} & \textbf{NS1} & \textbf{NS2} & \textbf{NS3} & \textbf{NS4} \\
\hline \hline
$\beta_0$ (int.)  	& -5.315  	&  -2.802	& -1.138 	& -2.916	& -1.167  \\ \hline
$\beta_1$ ($x$-coord) 	& -0.067 	& -0.037 	& -0.028 	& -0.039	&  -0.029 \\ \hline
$\beta_2$ ($y$-coord)  	& 0.034 	&  0.055 	& 0.0038 	& 0.053 	&  0.038 \\ \hline
$\lambda_1$  		& 2.772	& SV 	& SV 	& SV 	& SV \\ \hline
$\lambda_2$  		& 6.385 	& SV 	& SV 	& SV 	& SV \\ \hline
$\eta$  			& 0.336 	& SV 	& SV 	& SV 	& SV \\ \hline
$\tau^2$  			& 0.011 	& 0.005 	& 0.013 	& SV 	& SV \\ \hline
$\sigma^2$  		& 0.379 	& 0.289 	& SV 	& 0.275	& SV \\ \hline
\end{tabular}
\caption{ Parameter estimates for the five best spatial models fit to the precipitation data set, indicating which parameters are spatially-varying (SV). Recall that with an exponential correlation structure, the smoothness $\kappa$ is fixed to be $0.5$ for all models.}
\label{param_estimates}
\end{center}
\end{table}

\begin{figure}[!t]
\centering
\includegraphics[width=\textwidth]{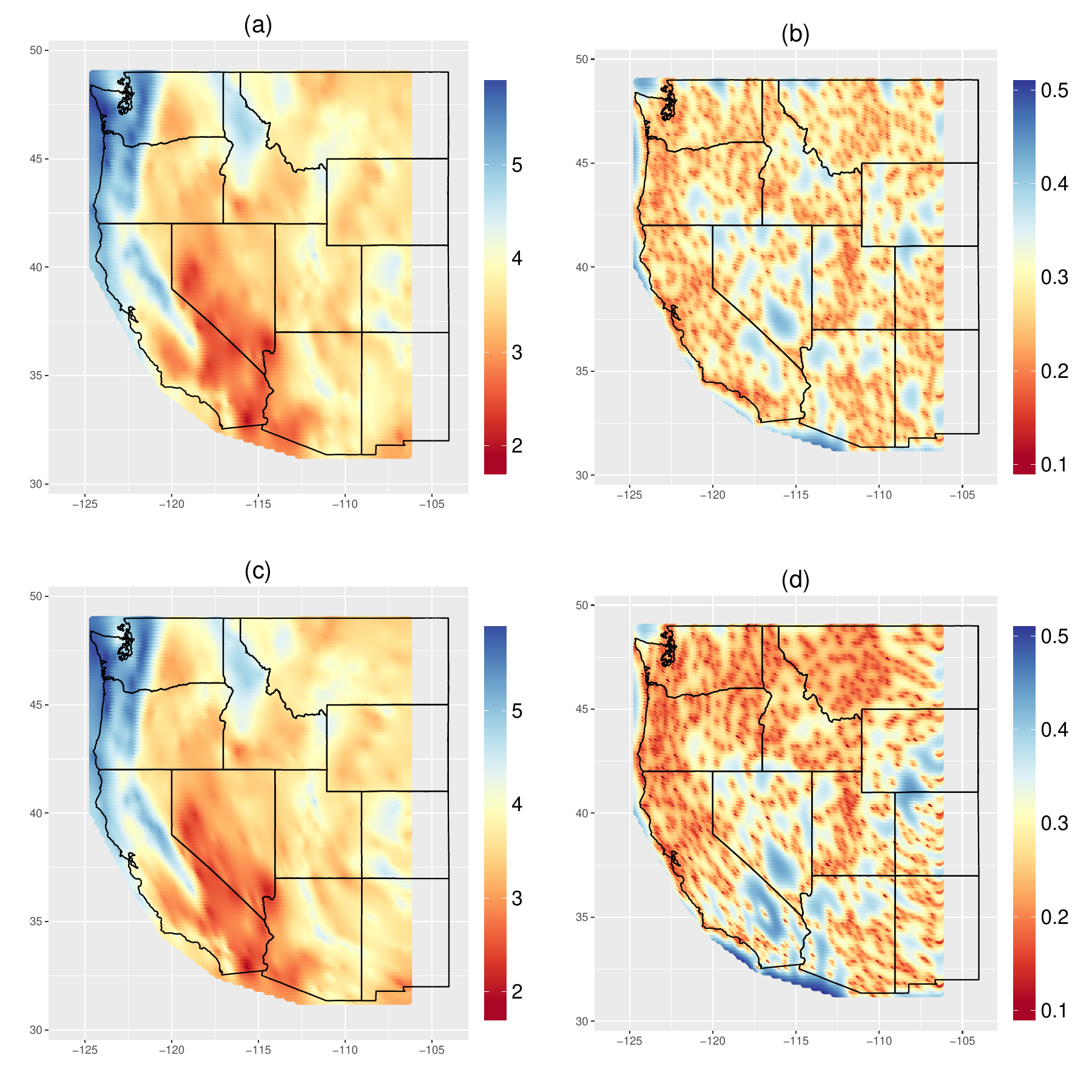}
\caption{ Predictions and prediction standard errors for the stationary model (plots (a) and (b)) and the nonstationary model NS1 (plots (c) and (d)). }
\label{precip_preds}
\end{figure}

While the gains are modest, all of the nonstationary models outperform the stationary model in terms of both MSPE and CRPS; the best model in terms of both MSPE and CRPS is NS1. The interpolated prediction maps and corresponding standard errors for the stationary model and NS1 are given in Figure \ref{precip_preds}. Note that the nonstationary model estimates the prediction errors much more flexibly, with larger errors in the far southwest (southern California, southern Nevada, and western Arizona) and smaller errors in the northern portion of the domain. The stationary model prediction errors are much more homogeneous, being highly dependent on the proximity of neighboring observations.

\begin{figure}[!t]
\centering
\includegraphics[trim={32 35 30 20mm}, clip, width=0.60\textwidth]{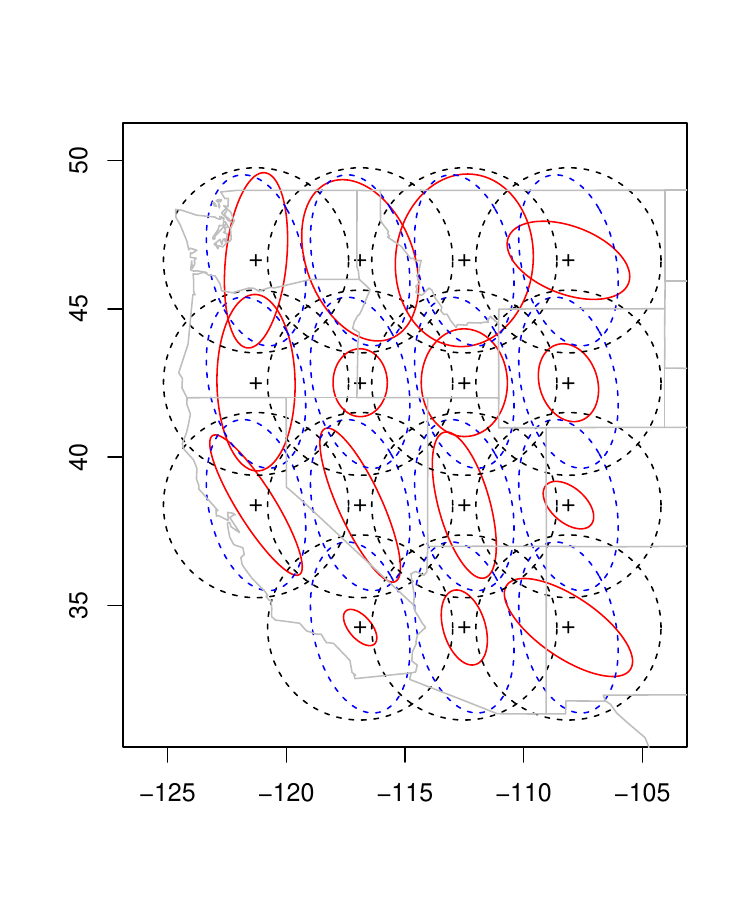}
\caption{ Estimated mixture component ellipses for the nonstationary model (red), the stationary (anisotropic) model (blue), and the estimation region (dashed black). }
\label{precip_ellip}
\end{figure}

\begin{figure}[!t]
\includegraphics[width=\textwidth]{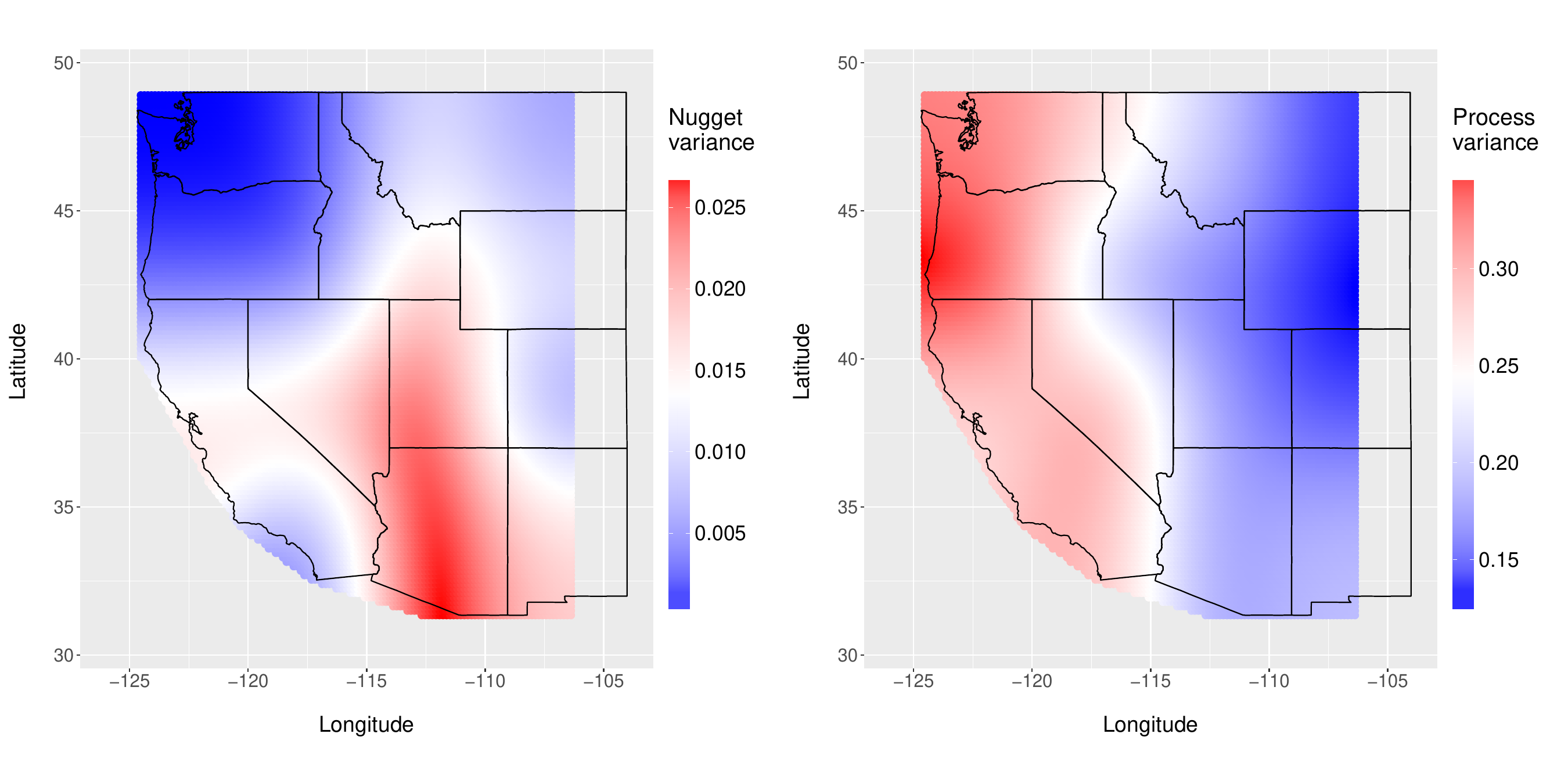}
\caption{ Plots of the estimated spatially-varying process variance $\sigma^2$ (right) and nugget variance $\tau^2$ (left) for model NS4. }
\label{precip_NuggVar}
\end{figure}

\subsection{Visualizations of nonstationarity}

Graphical summaries can be made to visualize the spatially-varying parameter estimates not explicitly provided in Table \ref{param_estimates}. The fully nonstationary model NS4 will be used for the following visualizations, as all of its variance/covariance parameters are spatially-varying.

First, consider the locally-estimated anisotropy ellipses for each of the mixture component locations, shown by the solid red ellipses in Figure \ref{precip_ellip}. This plot also contains the corresponding ellipse from the stationary model (dashed blue), which is constant over space, and the neighborhood size (dotted gray) defined by the fit radius $r$. It is quite clear that precipitation in the western United States displays nonstationary behavior with respect to the anisotropy ellipses, as the local estimates are highly variable across the spatial domain. 

\begin{figure}[!t]
\includegraphics[width=\textwidth]{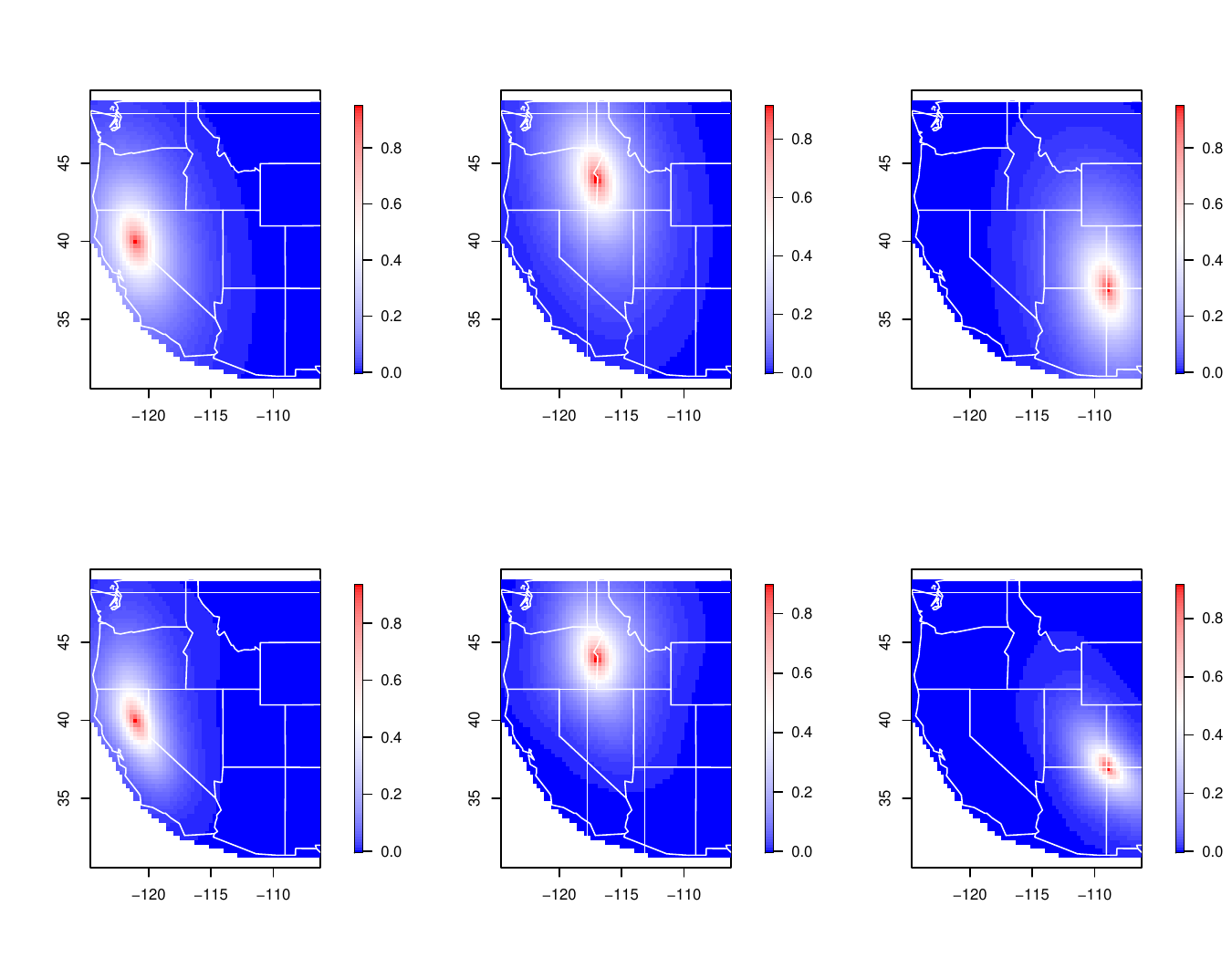}
\caption{ Correlation plots for three reference points, comparing the stationary model S (top) and nonstationary model NS4 (bottom). }
\label{precip_corrplots}
\end{figure}

Next, consider the estimates of the spatially-varying nugget variance ($\tau^2$) and process variance ($\sigma^2$) over the spatial region for NS4, shown in Figure \ref{precip_NuggVar}. The nugget variance is highly variable across the spatial domain, which may reflect variability in the monitoring devices. The process variance is largest along the coastal United States, which is intuitive based on its proximity to the ocean and highly variable topography. In general, areas with larger process variance correspond to smaller nugget variance, although this is not true everywhere (e.g., central Montana).

Finally, estimated correlation plots for three reference points are given in Figure \ref{precip_corrplots}. This plot nicely illustrates the fact that the nonstationary model allows the spatial dependence structure to change over space, while the stationary model estimates a constant correlation structure. 
In addition to allowing the orientation of the anisotropy ellipses to change over space (as seen in Figure \ref{precip_ellip}), the nonstationary model allows for non-elliptical correlation patterns.

\section{Discussion} \label{discussion}

In this paper, we have presented a nonstationary spatial Gaussian process model that is highly flexible yet amenable to computationally efficient inference, as shown through its implementation in the new \pkg{convoSPAT} package for \proglang{R}. In fact, for a moderately large real data set, fitting the nonstationary model is significantly faster than fitting the stationary model (at least using \code{Aniso_fit}; see Table \ref{realdata_summary}). The model also allows for visualization of the estimated nonstationarity, for both the spatially-varying variance parameters and correlation structure. 

The tradeoff for the computational tractability of this model is that uncertainty in the locally-estimated parameters is not accounted for in global estimation and, more seriously, this uncertainty is not quantified in the parameter estimation. Furthermore, the model is not completely pre-packaged, in that the user must specify the mixture component grid, fit radius, and tuning parameter, and the resulting parameter estimates and predictions may be sensitive to these choices. Since uncertainty in parameter estimates is not provided, it is difficult to determine if a nonstationary model is needed or if a stationary model would be sufficient. However, the primary goal of this model is to provide a ``quick and dirty'' way to fit a nonstationary Gaussian process model to spatial data, allowing new nonstationary methods to be compared with another model that is also nonstationary, instead of simply a stationary model. And, since the model can be fit very quickly, a practitioner can fit the model for many choices of the mixture component grid, fit radius, and tuning parameter (similar to the strategy in Section \ref{modelSumm}) and use the provided evaluation criteria (or other criteria) to choose a final model.

Finally, while this model provides a fast approach for modeling nonstationary spatial Gaussian processes, note that the same model could gain an additional degree of computational efficiency by parallelizing the LLE component of the model. Currently, the local parameters are estimated sequentially; however, the optimization for each mixture component location does not depend on the optimization for the other mixture component locations. Thus, a more efficient algorithm might estimate all of these components simultaneously, greatly reducing the overall computation time even further. A future version of \pkg{convoSPAT} will explore this possibility.



\bibliography{JSS_bib}

\end{document}